\documentclass[12pt, a4paper]{article}
\pdfoutput=1
\usepackage{graphicx}
\usepackage{amssymb}
\usepackage{amsmath}
\usepackage{bm}
\usepackage[table,dvipsnames]{xcolor} 
\usepackage{cite}
\usepackage{slashed}
\usepackage{epstopdf}            
\usepackage{epsfig}
\usepackage{here}
\usepackage{comment}
\usepackage{booktabs} 
\usepackage{colortbl} 
\usepackage{wrapfig}
\usepackage{ascmac}
\usepackage{fancybox}  
\usepackage{soul} 
\usepackage{adjustbox}

\allowdisplaybreaks
\renewcommand{\thefootnote}{\fnsymbol{footnote}}
\numberwithin{equation}{section} 
\def\beq#1\eeq{\begin{align}#1\end{align}}

\RequirePackage{xspace}
\usepackage{caption}
\definecolor{BlueViolet}{rgb}{0.2, 0.00, 0.7}
\definecolor{Blue}{rgb}{0.15, 0.00, 0.9}
\definecolor{light_blue}{rgb}{0.15, 0.35, 0.95}
\definecolor{kit_green}{rgb}{0
, 0.58823 
, 0.50980 
}
\usepackage[
colorlinks=true, linkcolor=light_blue,citecolor=light_blue,urlcolor=kit_green]{hyperref} 

\begin{document}
\sloppy 
\begin{titlepage}
\begin{center}
\hfill{KEK--TH--2744}\\
\vskip .3in

{\Large{\bf $b \to c$ semileptonic sum rule: \\ Current status and prospects}}\\
\vskip .3in

\makeatletter\g@addto@macro\bfseries{\boldmath}\makeatother

{ 
Motoi Endo$^{\rm (a,b,c)}$,
Syuhei Iguro$^{\rm (d,c)}$, 
Satoshi Mishima$^{\rm (e)}$, 
Ryoutaro Watanabe$^{\rm (f)}$
}
\vskip .3in
$^{\rm (a)}${\it KEK Theory Center, IPNS, KEK, Tsukuba 305--0801, Japan}\\\vspace{4pt}
$^{\rm (b)}${\it Graduate Institute for Advanced Studies, SOKENDAI, Tsukuba,\\ Ibaraki 305--0801, Japan} \\\vspace{4pt}
$^{\rm (c)}${\it Kobayashi-Maskawa Institute (KMI) for the Origin of Particles and the Universe, Nagoya University, Nagoya 464--8602, Japan}\\\vspace{4pt}
$^{\rm (d)}${\it Institute for Advanced Research (IAR), Nagoya University,\\ Nagoya 464--8601, Japan}\\\vspace{4pt}
$^{\rm (e)}${\it Department of Liberal Arts, Saitama Medical University, Moroyama,\\ Saitama 350-0495, Japan}\\\vspace{4pt}
$^{\rm (f)}${\it Institute of Particle Physics and Key Laboratory of Quark and Lepton Physics (MOE), Central China Normal University, Wuhan, Hubei 430079, China}
\end{center}
\vskip .15in

\begin{abstract}
The $b \to c$ semileptonic sum rules provide relations between the decay rates of $B \to D^{(*)} \tau\bar\nu$ and $\Lambda_b \to \Lambda_c \tau\bar\nu$. 
Starting from the heavy quark and zero-recoil limits, we revisit the derivation of the sum rule for total decay rates.
We then examine deviations from the limits and investigate corrections arising from realistic hadron masses and higher-order contributions to form factors, taking account of uncertainties.
We show that these corrections are negligible compared to current experimental uncertainties, indicating that the sum rule is useful for cross-checking experimental consistency and testing the validity of the Standard Model predictions.
In future, precise determinations of the form factors particularly for the tensor operator will be necessary to compare the sum rule predictions with $\Lambda_b \to \Lambda_c \tau\bar\nu$ data from the LHCb experiment and the Tera-Z projects.
\end{abstract}
{\sc ~~~~ Keywords: Heavy quark symmetry, $b \to c$ semileptonic sum rule} 
\end{titlepage}

\setcounter{page}{1}
\renewcommand{\thefootnote}{\#\arabic{footnote}}
\setcounter{footnote}{0}

\hrule
\tableofcontents
\vskip .2in
\hrule
\vskip .4in


\section{Introduction}
\label{sec:intro}

Lepton-flavor universality is a key prediction of the Standard Model (SM).
Measurements of $R_{D^{(*)}} ={\rm{BR}}(B\to D^{(*)} \tau\bar\nu)/{\rm{BR}}(B\to D^{(*)} \ell\bar\nu)$ with $\ell = e,\mu$ have shown deviations from the SM predictions at the 3.8$\sigma$ level~\cite{HeavyFlavorAveragingSpring2025}.
These results may suggest contributions from new physics (NP) leading to an excess of the tau lepton modes. 
In contrast, the recent measurement of $R_{\Lambda_c} ={\rm{BR}}(\Lambda_b\to\Lambda_c\tau\bar\nu)/{\rm{BR}}(\Lambda_b\to\Lambda_c\ell\bar\nu)$~\cite{LHCb:2022piu} is consistent with the SM prediction, and it indicates a slight deficit of the tau lepton mode.
Such a situation motivates investigations into potential shortcomings in the experimental data and SM predictions. 

One of the non-trivial tests is provided by the $b \to c$ semileptonic sum rule~\cite{Blanke:2018yud,Blanke:2019qrx,Fedele:2022iib,Duan:2024ayo},
\begin{align}
 \frac{R_{\Lambda_c}}{R_{\Lambda_c}^{SM}} - \alpha_R \frac{R_{D}}{R_{D}^{SM}} - \beta_R \frac{R_{D^*}}{R_{D^*}^{SM}}=\delta_R \,.
 \label{eq:RSR}
\end{align}
The denominators denote the SM predictions, while the numerators may include NP contributions. 
The coefficients, $\alpha_R \sim 1/4$ and $\beta_R \sim 3/4$, are independent of NP effects. 
The NP dependencies arise through the correction term $\delta_R$, and the sum rule provides a powerful tool for cross-checking the consistency of experimental data and testing the validity of SM predictions, especially when $\delta_R$ is sufficiently small.

The sum rule \eqref{eq:RSR} was proposed initially based on empirical observations~\cite{Blanke:2018yud} and investigated further in Refs.~\cite{Blanke:2019qrx,Fedele:2022iib,Duan:2024ayo}. 
Subsequently, another relation involving differential decay rates was derived within the heavy quark limit using the heavy quark effective theory (HQET)~\cite{Endo:2025fke},\footnote{
Sum rules are also obtained for angular observables~\cite{Endo:2025cvu}.
In this paper, we are interested in the sum rule for the total decay rates and that for the differential decay rates with respect to $w$.
}
\begin{align}
 \frac{\kappa_{\Lambda_c}}{\zeta(w)^2}-\frac{2}{1+w} \frac{\kappa_{D}+\kappa_{D^*}}{\xi(w)^2} = 0 \,,
 \label{eq:DDRSR}
\end{align}
where $\zeta(w)$ and $\xi(w)$ denote the leading-order Isgur-Wise (IW) functions for ground-state baryons and mesons, respectively.
Here, $\kappa_{H_c}^{w}=d\Gamma^{H_c}/dw$ represents the differential decay rates with $\Gamma^{H_c}=\Gamma(H_b\to H_c \tau\bar\nu)$ and $w = (m_{H_b}^2+m_{H_c}^2-q^2)/(2m_{H_b}m_{H_c})$.
The variable $q^2$ denotes the invariant mass of the leptons.
The sum rule \eqref{eq:DDRSR} is subject to deviations due to realistic hadron masses and higher-order contributions to form factors, reflecting the finite heavy quark masses.
Additionally, phase-space integrals over the differential decay rates contribute to $\delta_R$ when connecting Eq.~\eqref{eq:DDRSR} to Eq.~\eqref{eq:RSR}.
However, it has not yet been clarified why the coefficients take values around $\alpha_R = 1/4$ and $\beta_R = 3/4$.
Furthermore, although $\delta_R$ was discussed in Ref.~\cite{Endo:2025fke}, its uncertainty was neglected.
It is worth investigating whether the current experimental data and SM predictions are consistent with the sum rule \eqref{eq:RSR}.

In this paper, we revisit the $b \to c$ semileptonic sum rules.
We begin by outlining the derivation of Eq.~\eqref{eq:RSR} from Eq.~\eqref{eq:DDRSR} and clarify how the coefficients $\alpha_R = 1/4$ and $\beta_R = 3/4$ are obtained.
Then, we take both experimental and theoretical uncertainties into account.
In particular, the theoretical uncertainties mainly originate from the hadronic form factors.
Currently, two approaches are often employed to parametrize the form factors, namely those based on HQET~\cite{Isgur:1989vq,Neubert:1993mb} and the Boyd-Grinstein-Lebed (BGL) parameterization~\cite{Boyd:1995cf}.
We analyze the sum rules based on both frameworks and study the discrepancies between them.
Finally, we apply the sum rules to the current experimental data and discuss future prospects.

Theoretical uncertainties of $\delta_R$ arising from form factors have been studied in specific cases.
Reference~\cite{Duan:2024ayo} employed the BGL approach for both $B \to D^{(*)}$ and $\Lambda_b \to \Lambda_c$ transitions.
Although HQET-based form factors are currently available~\cite{Bernlochner:2017jka,Iguro:2020cpg,Bernlochner:2018kxh,Bernlochner:2018bfn}, they have not been applied to the sum rule, particularly for the $\Lambda_b \to \Lambda_c$ form factors.\footnote{Reference~\cite{Iguro:2024hyk} adopted HQET to obtain the $B \to D^{(*)}$ form factors, while the BGL parameterization was used for $\Lambda_b \to \Lambda_c$.
Moreover, the uncertainties were not investigated.}
In particular, the BGL form factors often involve large uncertainties. 
Although the probability distributions of $\delta_R$ have been assumed to be Gaussian in the literature, this approximation does not always hold when the uncertainties are large.
Therefore, we construct probability distribution functions using toy Monte Carlo simulations to evaluate the theoretical uncertainties. 

The rest of the paper is organized as follows.
In Sec.~\ref{sec:model}, we introduce the effective operators describing NP contributions. 
In Sec.~\ref{sec:SR_total}, the $b \to c$ semileptonic sum rule for the total decay rate is derived starting from Eq.~\eqref{eq:DDRSR}.
Then, we study the current status and prospects of the sum rule, taking account of theoretical and experimental uncertainties.
Sec.~\ref{sec:conclusion} is devoted to conclude our findings and discussion.
In Appendix~\ref{sec:figures}, we show additional figures. 
We also study the sum rule for the differential decay rate in Appendix~\ref{sec:SR_differential}.

\section{New physics framework}
\label{sec:model}

In this paper, we assume that NP contributes only to the $b\to c \tau\bar\nu$ transitions. 
The weak effective Hamiltonian is introduced as 
\begin{align}
 \label{eq:Hamiltonian}
 {\cal {H}}_{\rm{eff}}= 2 \sqrt2 \, G_FV_{cb}\biggl[ (1+C_{V_L})O_{V_L}+C_{S_L}O_{S_L}+C_{S_
R}O_{S_R}+C_{T}O_{T}\biggl]\,.
\end{align}
We consider dimension-six effective operators given by
\begin{align}
 &O_{V_L} = (\overline{c} \gamma^\mu P_Lb)(\overline{\tau} \gamma_\mu P_L \nu_{\tau})\,,\,\,\, 
 O_{S_R} = (\overline{c}  P_Rb)(\overline{\tau} P_L \nu_{\tau})\,, \label{eq:operator}\nonumber \\
 &O_{S_L} = (\overline{c} P_L b)(\overline{\tau} P_L \nu_{\tau})\,,\,\,\, O_{T} = (\overline{c}  \sigma^{\mu\nu}P_Lb)(\overline{\tau} \sigma_{\mu\nu} P_L \nu_{\tau}) \,,
\end{align}
where $P_{L(R)}=(1\mp\gamma_5)/2$ is a chirality projection operator. 
The NP contribution is encoded in the Wilson coefficients (WCs) of $C_X$, which are normalized with the SM factor, $2 \sqrt2 G_FV_{cb}$. 
In this framework, the SM limit corresponds to $C_{X} = 0$ for $X=V_{L}$, $S_{L,R}$, and $T$. 
We also assume that the neutrinos in Eq.~\eqref{eq:Hamiltonian} are left-handed.

\section{Sum rule for total decay rate}
\label{sec:SR_total}

\subsection{Formulation}
\label{sec:formula_total}

In this section, we derive the $b \to c$ semileptonic sum rule for the total decay rate, starting from Eq.~\eqref{eq:DDRSR}.\footnote{
The analysis of the $b \to c$ semileptonic sum rule for differential decay rates is presented in Appendix~\ref{sec:SR_differential}.
}
According to Ref.~\cite{Endo:2025fke}, the relation of Eq.~\eqref{eq:DDRSR} holds exactly for any NP contributions described by $C_X$ in the heavy quark limit.
In this limit, the hadron masses satisfy $m_B = m_{\Lambda_b} = m_b$ and $m_D = m_{D^*} = m_{\Lambda_c} = m_c$, where $m_{b,c}$ are the bottom and charm quark masses, respectively.
Besides the IW function is approximated with the leading-order IW function. 
By integrating both sides of Eq.~\eqref{eq:DDRSR} over the phase space, the relation becomes
\begin{align}
 \int_{1}^{w_{\rm max}}\!\!\!\!\!\!dw\, \frac{\kappa_{\Lambda_c}}{\zeta(w)^2} = 
 \int_{1}^{w_{\rm max}}\!\!\!\!\!\!dw\, \frac{2}{1+w} \frac{\kappa_{D} + \kappa_{D^*}}{\xi(w)^2} \,,
 \label{eq:DDRSR_int}
\end{align}
where the maximal value of $w$ is given by
\begin{align}
 w_{\rm max} = \frac{m_b^2+m_c^2-m_\tau^2}{2m_bm_c} \,.
\end{align}
Next, we additionally consider the zero-recoil limit, $m_c \to m_b$ and $m_\tau/m_b \to 0$, leading to $w_{\rm max} \to 1$.\footnote{
Such a combination of the approximations corresponds to the small-velocity (Shifman-Voloshin) limit, $m_b+m_c \gg m_b-m_c \gg \bar\Lambda$, where $\bar\Lambda$ is a QCD scale.
See, {\it e.g.,} Ref.~\cite{Buchalla:2002pd}.
}
The differential decay rates $\kappa_{H_c}$ approach zero, while the IW functions are approximated by $\xi(w) \to 1$ and $\zeta(w) \to 1$.
Hence, Eq.~\eqref{eq:DDRSR_int} is simplified as
\begin{align}
&
 \int_{1}^{w_{\rm max}}\!\!\!\!\!\!dw\, \frac{\kappa_{\Lambda_c}}{\zeta(w)^2}
 \to \int_{1}^{w_{\rm max}}\!\!\!\!\!\!dw\, \kappa_{\Lambda_c} 
 = \Gamma_{\Lambda_c} \,, \\
&
 \int_{1}^{w_{\rm max}}\!\!\!\!\!\!dw\, \frac{2}{1+w} \frac{\kappa_{D} + \kappa_{D^*}}{\xi(w)^2}
 \to \int_{1}^{w_{\rm max}}\!\!\!\!\!\!dw\, (\kappa_{D} + \kappa_{D^*})
 = \Gamma_{D} + \Gamma_{D^*} \,.
\end{align}
As we will see later in Eqs.~\eqref{eq:total_SM_1} and \eqref{eq:total_SM_2}, the non-zero contributions start from $\epsilon^5$ where $\epsilon = 1 - m_c/m_b$ is defined.
Consequently, we obtain the following relation among the total decay rates,
\begin{align}
 \Gamma_{\Lambda_c} = \Gamma_{D} + \Gamma_{D^*}\,.
 \label{eq:gamma_1}
\end{align}
This holds for any NP contribution, $C_X$, in the heavy quark and zero-recoil limits.\footnote{
Equation \eqref{eq:gamma_1} has been known for the SM contribution, $\Gamma(H_b \to H_c \ell\bar\nu)^{\rm SM}$, where $\ell$ denotes light leptons. 
See, {\it e.g.}, Ref.~\cite{Boyd:1995ht}.
Also, by using the experimental result of $\Gamma_{\Lambda_c}$ in Ref.~\cite{Bernlochner:2022hyz}, the experimental data do not satisfy Eq.~\eqref{eq:gamma_1}.
Such a discrepancy could originate from the experimental data and the violation of the limits.
}

Within those limits, we can find another relation among the SM contributions,
\begin{align}
 \Gamma_{D^*}^{\rm SM} = 3\, \Gamma_{D}^{\rm SM} \,.
 \label{eq:gamma_2}
\end{align}
This is not derived from Eq.~\eqref{eq:DDRSR}, but can be checked by explicitly integrating $\kappa_{D}$ and $\kappa_{D^*}$ over the phase space.\footnote{Equation \eqref{eq:gamma_2} has been known for semileptonic decay rates into light leptons, {\it i.e.,} $\eta \to 0$ in Eqs.~\eqref{eq:total_SM_1} and \eqref{eq:total_SM_2}. See, {\it e.g.,} Ref.~\cite{Buchalla:2002pd}.}
Within the heavy quark limit, the analytic forms of $\kappa_{D}$ and $\kappa_{D^*}$ are given by~\cite{Endo:2025cvu}
\begin{align}
 \kappa_D^{\rm SM} &= 
 \mathcal{N} \,
 \bigg[ \left( 1 + \frac{1}{2} \rho^2 \right) \frac{(1+r)^2 (w^2-1)}{\hat q^2}
 + \frac{3}{2} \rho^2 \frac{(1-r)^2(w+1)^2}{\hat q^2} \bigg] 
 \, \xi(w)^2 \,, \\
 \kappa_{D^*}^{\rm SM} &=
 \mathcal{N} \,
 \bigg[ \left( 1 + \frac{1}{2} \rho^2 \right) 
 \bigg[ 4w(w+1) 
 + \frac{(1-r)^2(w+1)^2}{\hat q^2} \bigg] 
 + \frac{3}{2} \rho^2 \frac{(1+r)^2(w^2-1)}{\hat q^2} \bigg] 
 \, \xi(w)^2 \,,
\end{align}
where the prefactor is defined as
\begin{align}
 \mathcal{N} = \frac{G_F^2 |V_{cb}|^2 \eta_{\rm EW}^2 m_b^5}{48\pi^3} 
 \hat q^2 r^3 \sqrt{w^2-1} 
 \left( 1 - \rho^2 \right)^2 \,.
\end{align}
Here, $\hat q^2 = q^2/m_b^2 = 1 - 2rw + r^2$, $r = m_c/m_b$, and $\rho = m_\tau/\sqrt{q^2}$ are introduced.
Also, $\eta_{\rm EW}$ is an EW correction. 
By performing the phase space integration in the zero-recoil limit, the total decay rate for the SM contributions are obtained as
\begin{align}
 \Gamma_{D}^{\rm SM} &= 
 \frac{G_F^2 |V_{cb}|^2 \eta_{\rm EW}^2 m_b^5}{120\pi^3}
 \,\epsilon^5
 \left[\sqrt{1-\eta^2}(2-9\eta^2-8\eta^4) + \frac{15}{2}\eta^4\,{\rm log}\frac{1+\sqrt{1-\eta^2}}{1-\sqrt{1-\eta^2}}\right]\,, 
 \label{eq:total_SM_1} \\
 \Gamma_{D^*}^{\rm SM} &= 
 \frac{G_F^2 |V_{cb}|^2 \eta_{\rm EW}^2 m_b^5}{40\pi^3}
 \,\epsilon^5
 \left[\sqrt{1-\eta^2}(2-9\eta^2-8\eta^4) + \frac{15}{2}\eta^4\,{\rm log}\frac{1+\sqrt{1-\eta^2}}{1-\sqrt{1-\eta^2}}\right]\,.\,
 \label{eq:total_SM_2} 
\end{align}
where $\eta = m_\tau/(m_b-m_c)$ takes a non-zero value even in the zero-recoil limit.
These results satisfy Eq.~\eqref{eq:gamma_2}.
Equation \eqref{eq:gamma_2} holds only for the SM contributions and is violated in the presence of $C_X$ with $X=S_{L,R}$ and $T$.
By combining Eqs.~\eqref{eq:gamma_1} and \eqref{eq:gamma_2}, the SM contribution $\Gamma_{\Lambda_c}^{\rm SM}$ is given by 
\begin{align}
 \Gamma_{\Lambda_c}^{\rm SM} = \Gamma_{D}^{\rm SM} + \Gamma_{D^*}^{\rm SM} = 4\, \Gamma_{D}^{\rm SM} \,.
 \label{eq:gamma_3}
\end{align}

By normalizing both sides of Eq.~\eqref{eq:gamma_1} with $\Gamma_{\Lambda_c}^{\rm SM}$ and by using Eqs.~\eqref{eq:gamma_2} and \eqref{eq:gamma_3}, the sum rule for the total decay rates is derived as
\begin{align}
 \frac{\Gamma_{\Lambda_c}}{\Gamma_{\Lambda_c}^{\rm SM}} = 
 \frac{\Gamma_{D}^{\rm SM}}{\Gamma_{\Lambda_c}^{\rm SM}} \frac{\Gamma_{D}}{\Gamma_{D}^{\rm SM}} + \frac{\Gamma_{D^*}^{\rm SM}}{\Gamma_{\Lambda_c}^{\rm SM}} \frac{\Gamma_{D^*}}{\Gamma_{D^*}^{\rm SM}} =
 \frac{1}{4} \frac{\Gamma_{D}}{\Gamma_{D}^{\rm SM}} + \frac{3}{4} \frac{\Gamma_{D^*}}{\Gamma_{D^*}^{\rm SM}} \,.
 \label{eq:gamma_4}
\end{align}
Since we assume that NP contributes only to $b\to c \tau\bar\nu$, the light-lepton channels are governed by the SM, {\it i.e.,} $\Gamma_{H_c}^\ell = \Gamma_{H_c}^{\ell\,{\rm SM}}$ with $\Gamma_{H_c}^\ell = \Gamma(H_b \to H_c\ell\bar\nu)$. 
Hence, by normalizing Eq.~\eqref{eq:gamma_4} with the decay rates of the light-lepton channels, we obtain the $b \to c$ semileptonic sum rule,
\begin{align}
 \frac{R_{\Lambda_c}}{R_{\Lambda_c}^{SM}} = a^{\rm HQ} \frac{R_{D}}{R_{D}^{SM}} + b^{\rm HQ} \frac{R_{D^*}}{R_{D^*}^{SM}} \,,~~~
 \text{with}~~
 a^{\rm HQ} = \frac{1}{4},~
 b^{\rm HQ} = \frac{3}{4} \,.
 \label{eq:RSR_limit}
\end{align}
Compared with Eq.~\eqref{eq:RSR}, we see that the coefficients correspond to $\alpha_R = a^{\rm HQ}$ and $\beta_R = b^{\rm HQ}$, and there is no NP-dependent correction term, $\delta_R$. 
It is important to note that Eq.~\eqref{eq:RSR_limit} holds in the heavy quark and zero-recoil limits.
Namely, deviations from these limits give rise to the correction term $\delta_R$.

In reality, these limits are broken, and the equality of the sum rules is violated.
The correction arises because the decay rates are evaluated using physical hadron mass spectrum and the form factors are not approximated by the leading-order IW functions.\footnote{
See Ref.~\cite{Endo:2025fke} for details of each contribution to the correction.
For example, with physical hadron mass spectra, the range of the phase-space integration depends on the decay channel, and such a variance induces $\delta_{\Lambda_c}^{kl}$ in Eq.~\eqref{eq:delta_int_mod}.}
This correction can be decomposed in terms of the WCs. 
In each ratio appearing in Eq.~\eqref{eq:RSR_limit}, the NP contributions enter through the decay rate in the numerator of $R_{H_c}$.
Accordingly, $R_{H_c}$ can be expressed as
\begin{align}
 R_{H_c} \equiv \sum_{ij} \mathcal{C}_i \mathcal{C}_j^*\,R^{ij}_{H_c} \,,
\end{align}
where the Wilson coefficient factors $\mathcal{C}_i$ are defined as
\begin{align}
 \mathcal{C}_i = 
 \begin{cases}
 \, 1+C_{V_L} \\
 \, C_{S_L} \\
 \, C_{S_R} \\
 \, C_T
 \end{cases}
 \text{for}~~
 i=
 \begin{cases}
 \, V_L, \\
 \, S_L, \\
 \, S_R, \\
 \, T.
 \end{cases}
\end{align}
The SM prediction corresponds to $C_{X} = 0$, and thus, is given by $R^{\rm SM}_{H_c} = R^{V_LV_L}_{H_c}$.
Additionally, since $R^{S_LS_L}_{H_c} = R^{S_RS_R}_{H_c}$ is generally satisfied, we denote both $(ij) = (S_LS_L)$ and $(S_RS_R)$ as $(ij) = (SS)$.
Hence, the correction to the sum rule is expressed as
\begin{align}
 \delta_{\Lambda_c}^{\rm HQ} = \sum_{ij}  \mathcal{C}_i \mathcal{C}_j^* \,\delta_{\Lambda_c}^{\rm HQ}(ij)\,,
 ~~~
 \delta_{\Lambda_c}^{\rm HQ}(ij) \equiv 
 \frac{R^{ij}_{\Lambda_c}}{R^{\rm SM}_{\Lambda_c}} - a^{\rm HQ} \frac{R^{ij}_{D}}{R^{\rm SM}_{D}} - b^{\rm HQ} \frac{R^{ij}_{D^*}}{R^{\rm SM}_{D^*}} \,.
 \label{eq:delta_int}
\end{align}
Since $\delta_{\Lambda_c}^{\rm HQ}(V_LV_L) = 0$ is obviously satisfied, $\delta_{\Lambda_c}^{\rm HQ}$ consists solely of NP contributions.

There are additional degrees of freedom to modify the coefficients $a^{\rm HQ}$ and $b^{\rm HQ}$ when the heavy quark and zero-recoil limits are broken. 
In particular, we can suppress some of the $\mathcal{C}_i \mathcal{C}_j^*$ terms in Eq.~\eqref{eq:delta_int} by shifting them.\footnote{
Such a prescription was also adopted in Refs.~\cite{Blanke:2018yud,Blanke:2019qrx,Fedele:2022iib,Duan:2024ayo}.
}
Thus, instead of Eq.~\eqref{eq:delta_int}, we redefine the deviation as
\begin{align}
 \delta_{\Lambda_c}^{kl} = \sum_{ij}  \mathcal{C}_i \mathcal{C}_j^* \,\delta_{\Lambda_c}^{kl}(ij)\,,
 ~~~
 \delta_{\Lambda_c}^{kl}(ij) \equiv 
 \frac{R^{ij}_{\Lambda_c}}{R^{\rm SM}_{\Lambda_c}} - a^{kl} \frac{R^{ij}_{D}}{R^{\rm SM}_{D}} - b^{kl} \frac{R^{ij}_{D^*}}{R^{\rm SM}_{D^*}} \,,
 \label{eq:delta_int_mod}
\end{align}
When $\delta_{\Lambda_c}^{kl}(V_LV_L) = 0$ and $\delta_{\Lambda_c}^{kl}(kl) = 0$ are required, the coefficients are given by
\begin{align}
 a^{kl} = \frac{R_{\Lambda_c}^{kl}/R_{\Lambda_c}^{\rm SM}-R_{D^*}^{kl}/R_{D^*}^{\rm SM}}{R_{D}^{kl}/R_{D}^{\rm SM}-R_{D^*}^{kl}/R_{D^*}^{\rm SM}} \,,~~~
 b^{kl} = \frac{R_{D}^{kl}/R_{D}^{\rm SM}-R_{\Lambda_c}^{kl}/R_{\Lambda_c}^{\rm SM}}{R_{D}^{kl}/R_{D}^{\rm SM}-R_{D^*}^{kl}/R_{D^*}^{\rm SM}} \,.
 \label{eq:coefficient_int_mod}
\end{align}
One can check that $a^{kl} + b^{kl} = 1$ is satisfied, guaranteeing $\delta_{\Lambda_c}^{kl}(V_LV_L) = 0$.
The correction vanishes, {\it i.e.,} $\delta_{\Lambda_c}^{kl} \to 0$, if we take the heavy quark and zero-recoil limits in Eq.~\eqref{eq:delta_int_mod}, because $\delta_{\Lambda_c}^{kl}(ij)\neq 0$ arises due to the breaking of these limits. 
Additionally, the sum rule coefficients also become $a^{kl} \to 1/4$ and $b^{kl} \to 3/4$ in the limits.
It is also noted that $\delta_R$ in Eq.~\eqref{eq:RSR} depends on a choice of the coefficients of the sum rule, {\it i.e.,} $\delta_R  = \delta_{\Lambda_c}^{\rm HQ}$ or $\delta_{\Lambda_c}^{kl}$ depending on $a^{kl}$ and $b^{kl}$.

\subsection{Form factor}
\label{sec:ff}

In this section, we summarize the form factor parameterizations used in our analysis.
As mentioned above, $\delta_R$ arises due to the realistic hadron mass spectrum and higher-order corrections to the form factors.
Since the hadron masses have been measured precisely, the dominant theoretical uncertainties stem from the form factors.
In the following analysis, we examine two form factor parameterizations, namely the one based on HQET~\cite{Isgur:1989vq,Neubert:1993mb} and the other using the BGL parameterization~\cite{Boyd:1995cf}.

Within HQET, the form factors simplify significantly due to the heavy quark symmetry. 
In the heavy quark limit, they are governed by single leading-order IW functions, denoted by $\xi(w)$ for $B \to D^{(*)}$ transitions and by $\zeta(w)$ for $\Lambda_b \to \Lambda_c$.
However, due to the breaking of the limit, the form factors include corrections as
\begin{align}
 h_X(w) = \xi(w) \left[ \hat h_{X,0} + \frac{\alpha_s}{\pi} \delta \hat h_{X,\alpha_s} + \frac{\bar \Lambda}{2m_b} \delta \hat h_{X,m_b} + \frac{\bar \Lambda}{2m_c} \delta \hat h_{X,m_c} + \cdots \right] \,, 
 \label{eq:FF_HQET}
\end{align} 
where $\hat h_{X,0} = 1$ or $0$ corresponds to the leading-order contributions.
This form applies to mesonic transitions. 
For baryonic ones, $\xi(w)$ is replaced by $\zeta(w)$.
The higher-order contributions $\delta \hat h_{X}$ depend on higher-order IW functions.
The leading and higher-order IW functions are determined by global fits to lattice QCD results, experimental data, and theoretical models such as QCD sum rule and light-cone sum rule (LCSR).
They are parameterized as power series of the conformal variable $z$, defined as 
\begin{align}
 z = \frac{\sqrt{w+1}-\sqrt{2}}{\sqrt{w+1}+\sqrt{2}}.
\end{align}
This mapping of $q^2$ to $z$ helps to improve the convergence of the expansion.
In the following analysis, we adopt the procedure outlined in Refs.~\cite{Iguro:2020cpg,Endo:2025cvu}.
Corrections are included up to $\mathcal{O}(\alpha_s, \bar \Lambda/m_{b,c}, \bar \Lambda^2/m_{c}^2)$ for mesons, following Refs.~\cite{Bernlochner:2017jka,Iguro:2020cpg}, and up to $\mathcal{O}(\alpha_s, \bar \Lambda/m_{b,c}, \alpha_s\bar \Lambda/m_{b,c}, \bar \Lambda^2/m_{c}^2)$ for baryons, following Refs.~\cite{Bernlochner:2018kxh,Bernlochner:2018bfn}.
Here, $\bar \Lambda$ denotes a typical QCD scale.
Specifically, we adopt the fit results from the z210 scenario for $B\to D^{(*)}$~\cite{Iguro:2020cpg}, and those from the $\hat{b}_{1,2} \neq 0$ scenario for $\Lambda_b \to \Lambda_c$~\cite{Bernlochner:2018bfn}.
The central values, uncertainties, and correlation matrices are found in Ref.~\cite{Iguro:2020cpg} with a minor correction for $B\to D^{(*)}$ and in Ref.~\cite{Bernlochner:2018kxh} for $\Lambda_b \to \Lambda_c$.

The BGL parameterization is based on fundamental principles such as analyticity and unitarity, providing a model-independent framework to describe form factors.
In contrast to HQET, the series expansions are not applied via the IW functions, but the form factors themselves are expanded as power series of $z$.
The expansion coefficients are determined using inputs from lattice QCD calculations, experimental data, and theoretical constraints.
In the following analysis, we adopt the approach outlined in Ref.~\cite{Duan:2024ayo}.
The central values, uncertainties, and correlation matrices are found in Ref.~\cite{Cui:2023jiw} for $B \to D^{(*)}$ transitions except for the tensor operator, and those for the tensor are provided in Ref.~\cite{Gubernari:2018wyi}. 
For $\Lambda_b \to \Lambda_c$ transitions, the fit results for the operators except the tensor are taken from Ref.~\cite{Detmold:2015aaa}, and those for the tensor operator are given in Ref.~\cite{Datta:2017aue}.
It is stressed that the BGL form factors are semi-independent, namely, the vector and tensor form factors are determined by the inputs separately.
In contrast, those based on HQET are expressed by the common IW functions, and thus, are correlated with each other.\footnote{The scalar form factors are related to the vector ones by the equations of motion in both approaches. 
See, {\it e.g.,} Ref.~\cite{Bernlochner:2017jka} for HQET and Ref.~\cite{Bigi:2016mdz} for BGL.}

We compare the inputs for the HQET form factors in Ref.~\cite{Iguro:2020cpg} and those for BGL in Ref.~\cite{Duan:2024ayo}. 
The latter incorporates the lattice results for $B \to D^{*}$ by FNAL/MILC~\cite{FermilabLattice:2021cdg}, which became available after Ref.~\cite{Iguro:2020cpg} was published.
It also includes an updated analysis of LCSR~\cite{Cui:2023jiw}.
Although lattice results from HPQCD~\cite{Harrison:2023dzh} and JLQCD~\cite{Aoki:2023qpa} currently exist, they have not yet been incorporated into either analysis.\footnote{
See Ref.~\cite{Bordone:2024weh} for comparisons of different combinations of experimental and theoretical inputs.
}
New experimental results from Belle~\cite{Belle:2023bwv} and Belle II~\cite{Belle-II:2023okj} are also expected to be incorporated in future analyses.\footnote{
A recent result from BaBar~\cite{BaBar:2023kug} on $B \to D\ell\bar\nu$ does not provide sufficient information for theorists to determine the form factors.
}
On the other hand, there is no lattice result for the tensor operator in $B \to D$ currently. 
Although such a result is available for $B \to D^{*}$~\cite{Harrison:2023dzh}, it appeared after Ref.~\cite{Iguro:2020cpg} was published. 
Since the analysis relies on the heavy quark symmetry, it cannot be directly applied to the BGL framework. 
Although the tensor form factor is provided by LCSR~\cite{Gubernari:2018wyi}, the result is not accurate enough especially to determine the BGL form factors. 
Finally, for $\Lambda_b \to \Lambda_c$, lattice results exist for both the tensor~\cite{Datta:2017aue} and other form factors~\cite{Detmold:2015aaa}, while there are no corresponding LCSR results. 
These lattice inputs are used in both HQET and BGL form factors.
Reanalyses will be required, as the most up-to-date information has not yet been incorporated into either form factor analysis, and additional inputs such as lattice QCD calculations, experimental data, and theoretical inputs are expected to become available.

\subsection{Numerical result}
\label{sec:result_int}

In this section, we analyze the $b \to c$ semileptonic sum rules with the HQET and BGL form factors.
We study four possible combinations of form factor parameterizations:
\begin{itemize}
\item HQET for both $B \to D^{(*)}$ and $\Lambda_b \to \Lambda_c$ form factors.
\item BGL for both form factors.
\item HQET for $B \to D^{(*)}$ and BGL for $\Lambda_b \to \Lambda_c$.
\item BGL for $B \to D^{(*)}$ and HQET for $\Lambda_b \to \Lambda_c$.
\end{itemize}
Among them, the second scenario has been considered in Ref.~\cite{Duan:2024ayo}. 
Here, although the sum rule parameters and their uncertainties are evaluated, they are assumed to follow Gaussian probability distributions.
We will revisit this assumption in the following analysis.
The third scenario was adopted in Ref.~\cite{Iguro:2024hyk}, although the uncertainties were not evaluated. 
Reference~\cite{Endo:2025fke} revisited and compared the first and second scenarios, again without incorporating uncertainties.
There are no studies based on the fourth scenario. 
In the following analysis, we examine all four scenarios and compare the results.
We perform toy Monte Carlo simulations to evaluate theoretical uncertainties.
The parameters in the form factors are assumed to follow Gaussian probability distributions.
Then, the sum rule parameters are evaluated. 
We construct their probability distribution functions and determine corresponding confidence intervals.
 
\begin{figure}[!t]
\begin{center}
\includegraphics[width=0.4\linewidth]{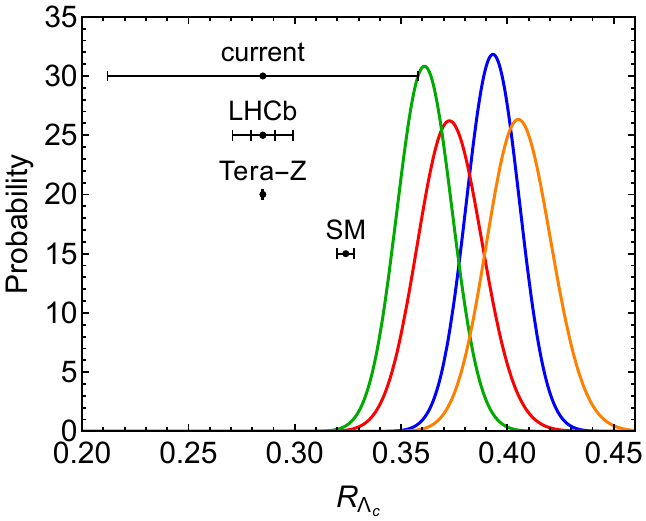}
\includegraphics[width=0.41\linewidth]{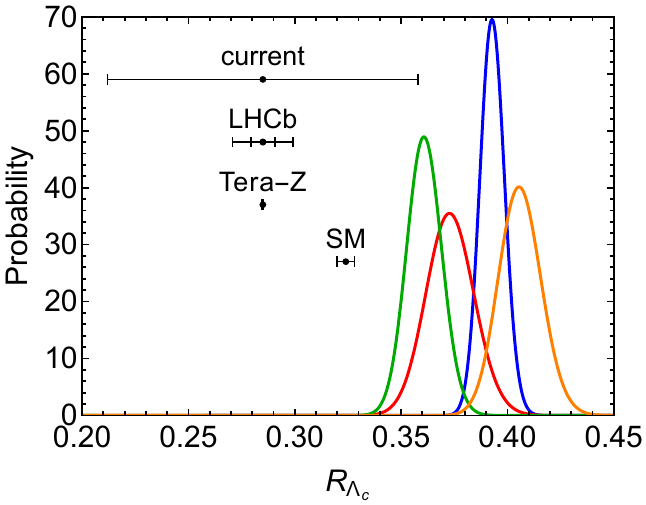}
\mbox{\raisebox{18mm}{\includegraphics[width=0.16\linewidth]{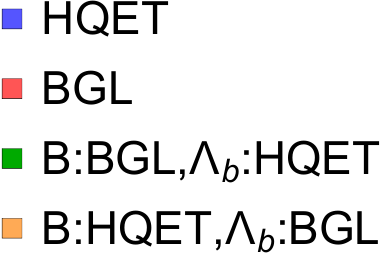}}}
\end{center}
\vspace{-.25cm}
\caption{
Sum rule predictions for $R_{\Lambda_c}$ using various form factors, assuming $a^{\rm HQ} = 1/4$, $b^{\rm HQ} = 3/4$, and $\delta_{\Lambda_c}^{\rm HQ} = 0$. 
For $R_D$ and $R_{D^*}$, the current experimental results~\cite{HeavyFlavorAveragingSpring2025} are used in the left panel, while future projections at Belle II with $\int\!\mathcal{L}\,dt=50\,{\rm ab}^{-1}$~\cite{Belle-II:2018jsg,Belle-II:2022cgf,ATLAS:2025lrr} are used in the right.
The data points for $R_{\Lambda_c}$ represent the current experimental result, future prospects at LHCb (pessimistic and optimistic scenarios), and projections at Tera-Z (with statistical uncertainties only), where the error bars indicate $68\,\%$ confidence intervals.
The SM prediction for $R_{\Lambda_c}$ is also displayed.
}
\label{fig:RLambda}
\end{figure}

In Fig.~\ref{fig:RLambda}, we show the probability distribution of $R_{\Lambda_c}$ predicted by the $b \to c$ semileptonic sum rule.
Here, we use Eq.~\eqref{eq:delta_int}, where the coefficients are given by $a^{\rm HQ} = 1/4$ and $b^{\rm HQ} = 3/4$, and we assume $\delta_{\Lambda_c}^{\rm HQ} = 0$.
In the sum rule, the SM values are evaluated using the above four combinations of form factor parameterizations.
Moreover, the current experimental values averaged by Ref.~\cite{HeavyFlavorAveragingSpring2025} are used for $R_D$ and $R_{D^*}$ in the left panel.
In the right, we adopt the prospect for $R_D$ and $R_{D^*}$ at the Belle II experiment with the integrated luminosity $\int\!\mathcal{L}\,dt=50\,{\rm ab}^{-1}$~\cite{Belle-II:2018jsg,Belle-II:2022cgf,ATLAS:2025lrr}, where the central values are unchanged from the current results. 

By comparing the four combinations of form factor parameterizations, we observe discrepancies among the predictions, though the tensions are not significant given the current theoretical uncertainties.  
The central value of $R_{\Lambda_c}$ is likely to be larger (smaller) when HQET (BGL) form factors are used for the $B \to D^{(*)}$ transitions.
In contrast, the discrepancies arising from the $\Lambda_b \to \Lambda_c$ form factor parameterization are less significant. 
On the other hand, the theoretical uncertainties associated with the form factors tend to be larger when the BGL is adopted, compared to the HQET-based parameterization.
In particular, the uncertainty in $R_{\Lambda_c}$ becomes larger when BGL is used for the $\Lambda_b \to \Lambda_c$ transition.
These discrepancies may be relaxed in future, especially through improvements in lattice QCD calculations, experimental data, and theoretical inputs.

\begin{figure}[!t]
\begin{center}
\includegraphics[width=0.41\linewidth]{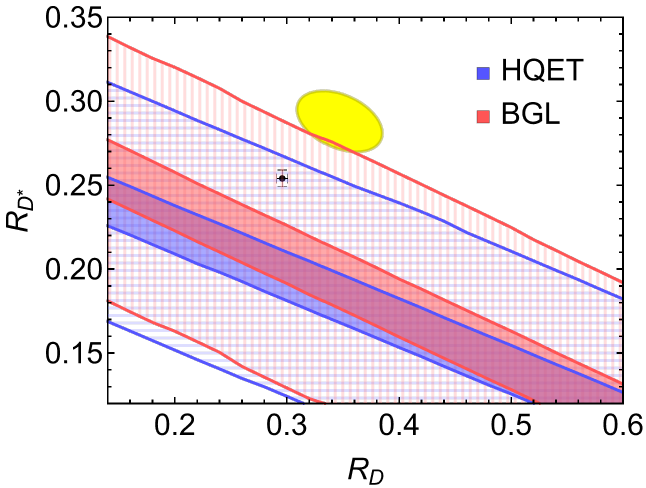}~~
\includegraphics[width=0.41\linewidth]{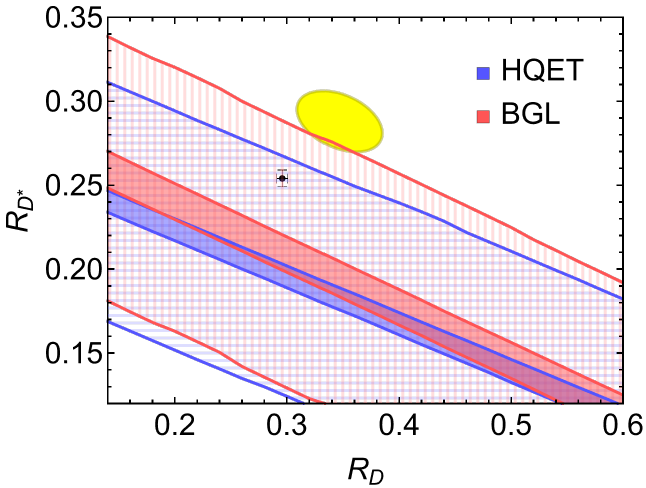}
\end{center}
\vspace{-.25cm}
\caption{
Sum rule predictions in the $R_D$--$R_{D^*}$ plane based on the current experimental value (hatched) and future projections (filled) for $R_{\Lambda_c}$.
The sum rule coefficients are taken as $a^{\rm HQ} = 1/4$ and $b^{\rm HQ} = 3/4$ with $\delta_{\Lambda_c}^{\rm HQ} = 0$.
The form factors are evaluated using HQET (blue) and BGL (red).
Each region corresponds to $68\,\%$ confidence level.
The experimental uncertainty in $R_{\Lambda_c}$ is expected to decrease according to the LHCb projections in the pessimistic (left) and optimistic (right) scenarios, while keeping the central value unchanged from the current result. 
The black point and yellow filled region respectively represent the SM prediction and the current experimental results for $R_D$ and $R_{D^*}$~\cite{HeavyFlavorAveragingSpring2025}.
}
\label{fig:RD_RDs}
\end{figure}

The Belle II experiment will improve measurements of $R_D$ and $R_{D^*}$.\footnote{
Although improvements are also expected from the LHCb experiment, its projected uncertainties are 2--3 times larger than the Belle II targets~\cite{ATLAS:2025lrr}. } 
Although the current uncertainty in the prediction of $R_{\Lambda_c}$ is dominated by the experimental results of $R_D$ and $R_{D^*}$, the uncertainties associated with the form factors are expected to become relevant in the future.
In particular, improvements of the $\Lambda_b \to \Lambda_c$ form factors are important especially for BGL to predict $R_{\Lambda_c}$ precisely. 

In Fig.~\ref{fig:RLambda}, we also present the current experimental results and future projections for $R_{\Lambda_c}$.
For the former, the decay rate of $\Lambda_b \to \Lambda_c \tau\bar\nu$ has been measured at the LHCb experiment~\cite{LHCb:2022piu}. 
If it is normalized with the SM prediction of $\Gamma(\Lambda_b \to \Lambda_c \mu\bar\nu)$, one obtains $R_{\Lambda_c} = (0.04/|V_{cb}|)^2 (0.285 \pm 0.073)$~\cite{Bernlochner:2022hyz}, which is shown in Fig.~\ref{fig:RLambda} using $|V_{cb}| = 0.04$.\footnote{
The decay rate of $\Lambda_b \to \Lambda_c \mu\bar\nu$ has been measured at DELPHI~\cite{DELPHI:2003qft}.
By combining its result with the LHCb value for $\Lambda_b \to \Lambda_c \tau\bar\nu$, one obtains $R_{\Lambda_c} = 0.242 \pm 0.076$~\cite{LHCb:2022piu}. 
}
It is noted that Ref.~\cite{Bernlochner:2022hyz} relies on HQET to evaluate the SM prediction. 
If instead BGL is adopted to the $\Lambda_b \to \Lambda_c$ form factors, the central value of $\Gamma(\Lambda_b \to \Lambda_c \mu\bar\nu)$ may shift by $\sim 1\sigma$. 
On the other hand, the LHCb prospects for $R_{\Lambda_c}$ with $\int\!\mathcal{L}\,dt=300\,{\rm fb}^{-1}$ are discussed in Ref.~\cite{Bernlochner:2021vlv}, where two scenarios are considered depending on the level of irreducible systematic uncertainty.
In the pessimistic scenario, a relative uncertainty of $5\%$ is assigned, whereas in the optimistic scenario, it is assumed to be $2\%$.
At the Tera-Z projects, a huge number of $\Lambda_b$ baryons are expected to be produced.
Although even more $\Lambda_b$ baryons may be produced at LHCb in the future, the decay could be measured more precisely at Tera-Z.
A relative precision in $R_{\Lambda_c}$ may reach as small as $9.8 \times 10^{-4}$~\cite{Ho:2022ipo,Ai:2024nmn}, although the systematic uncertainty is not included yet.
All these values are displayed by the black points with horizontal error bars in Fig.~\ref{fig:RLambda}.

\begin{table}[t]
\begin{center}
  \begin{tabular}{cccc} 
 Scenario & Parameter & Value &  Pull  \\ \hline
$S_L$ & $C_{S_L}$ & $-0.57\pm0.86\,i$ & 4.3\\
$S_R$ & $C_{S_R}$ & 0.18 & 3.9\\
$T$ & $C_{T}$ & $0.02 \pm 0.13\,i$ & 3.8\\
${\rm{R}}_2$ & $C_{S_L}=8.4\,C_T$ & $-0.09 \pm 0.56\,i$ & 4.4\\
${\rm{S}}_1$ & $C_{S_L}=-8.9\,C_T$ & $0.18$ & 4.1\\
${\rm{U}}_1$ & $C_{V_L}$,\,$\phi$ & $0.075,\,\pm 0.466\pi$ & 4.4\\ \hline
\end{tabular}
  \caption{Fit results for WCs in single-operator ($S_L$, $S_R$, $T$) and single leptoquark scenarios (${\rm R}_2$, ${\rm S}_1$, ${\rm U}_1$).
  The first column indicates the scenario, with the relevant WCs listed in the second column.
  For the ${\rm U}_1$ leptoquark, we consider $U(2)$-flavored scenario, which satisfies the relation $C_{S_R} = -3.7e^{i\phi}C_{V_L}$.
  See Ref.~\cite{Iguro:2024hyk} for further details.
  The best-fit values of the WCs at the $\mu_b$ scale are presented in the third column, and the fit quality is expressed by the pull value in the fourth column, whose definition is given in Ref.~\cite{Iguro:2024hyk}.\vspace{.15cm}\\
 }
  \label{Tab:scenario}
\end{center}   
\vspace{-.25cm}
\end{table}

\begin{figure}[!t]
\begin{center}
\includegraphics[width=0.37\linewidth]{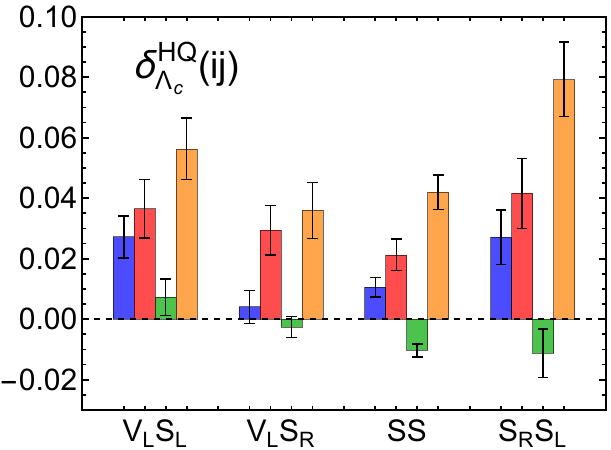}~
\includegraphics[width=0.227\linewidth]{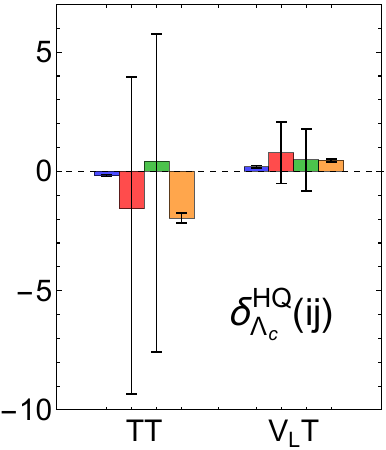}~
\includegraphics[width=0.365\linewidth]{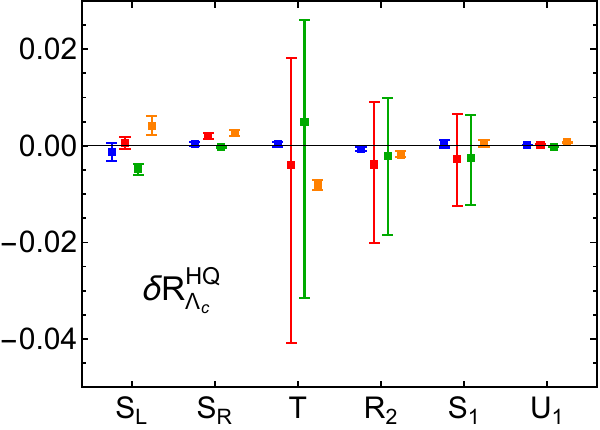}
\end{center}
\vspace{-.25cm}
\caption{
Correction to the sum rule for the total decay rates, $\delta^{\rm HQ}_{\Lambda_c}(ij)$ (left and center), and that to $R_{\Lambda_c}$ (right). 
The sum rule coefficients are assumed as $a^{\rm HQ}=1/4$ and $b^{\rm HQ}=3/4$.
Each bar represents the central value, and the error bar expresses its $68\,\%$ uncertainty.
The corrections vanish in the heavy quark and zero-recoil limits, shown by horizontal lines.
In the left and central panels, the horizontal items denote $(ij)$. 
In the right panel, the items correspond to the NP scenarios, whose WCs are given in Table~\ref{Tab:scenario}.
The color legend is the same as that in Fig.~\ref{fig:RLambda}.
}
\label{fig:delta_HQ}
\end{figure}

We also compare the sum rule prediction for $R_{\Lambda_c}$ with the SM prediction.
In Fig.~\ref{fig:RLambda}, we show $R_{\Lambda_c}^{\rm SM} = 0.324 \pm 0.004$, evaluated in Refs.~\cite{Bernlochner:2018kxh,Bernlochner:2018bfn} based on HQET.\footnote{
Reference~\cite{Duan:2024ayo} obtained $R_{\Lambda_c}^{\rm SM} = 0.332 \pm 0.010$ based on BGL.
} 
Since the current experimental results for $R_D$ and $R_{D^*}$ exceed their SM predictions, the sum rule prediction for $R_{\Lambda_c}$ becomes larger than $R_{\Lambda_c}^{\rm SM}$.
In contrast, the current experimental result for $R_{\Lambda_c}$ is slightly smaller than the SM prediction, though the tendency is not significant compared to the experimental uncertainty.
Nonetheless, due to the large experimental uncertainty in $R_{\Lambda_c}$, the sum rule prediction remains consistent with the current experimental result. 
Therefore, we find no shortcomings in the current experimental data and the SM predictions. 
This consistency can be further tested in future experiments, such as LHCb and Tera-Z.

Complementary to Fig.~\ref{fig:RLambda}, we present, in Fig.~\ref{fig:RD_RDs}, the sum rule prediction on the $R_D$--$R_{D^*}$ plane using the experimental value and future projections for $R_{\Lambda_c}$.
Currently, the sum rule prediction is consistent with both the experimental results and the SM predictions for $R_D$ and $R_{D^*}$.
As mentioned above, LHCb is expected to reduce the experimental uncertainties in $R_{\Lambda_c}$. 
In the plot, we show both pessimistic and optimistic scenarios regarding the irreducible systematic uncertainties, which are explained above.
Although Tera-Z may achieve higher precision, the sum rule prediction would remain nearly identical to the LHCb optimistic scenario, since the uncertainty is dominated by the form factors.
If this uncertainty would be reduced, Tera-Z could provide a better prediction on the $R_D$--$R_{D^*}$ plane.

\begin{figure}[!t]
\begin{center}
\includegraphics[width=0.4\linewidth]{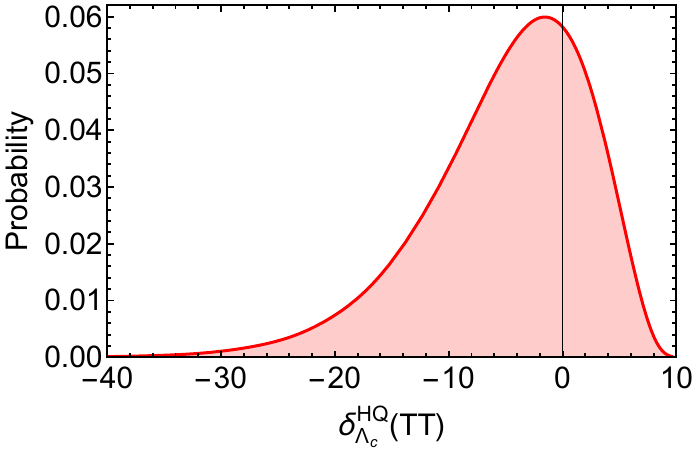}
\end{center}
\vspace{-.25cm}
\caption{
Probability distributions for $\delta^{\rm HQ}_{\Lambda_c}(TT)$ using the BGL form factors.
\vspace{0.1cm}\\
}
\label{fig:prob_TT}
\end{figure}

So far, we have assumed that $\delta_{\Lambda_c}$ is sufficiently small. 
However, it may become non-negligible if the NP contributions $C_X$ are sizable. 
In Fig.~\ref{fig:delta_HQ}, we present the potential size of $\delta^{\rm HQ}_{\Lambda_c}(ij)$ for $(ij) = (V_LS_L)$, $(V_LS_R)$, $(SS)$, $(S_RS_L)$, $(TT)$, and $(TV_L)$ in the left and central panels.
The error bars represent $68\,\%$ probability intervals, obtained by constructing probability distributions using toy Monte Carlo simulations.
We observe that when the $B \to D^{(*)}$ form factors are evaluated using the BGL parameterization, the uncertainties associated with the tensor operator become large and asymmetric around the central values, particularly for $(ij) = (TT)$.
Such a behavior is clearly seen in Fig.~\ref{fig:prob_TT}, where we show the probability distribution for $\delta^{\rm HQ}_{\Lambda_c}(TT)$. 
In contrast, the probability distributions for other $\delta^{\rm HQ}_{\Lambda_c}(ij)$ can be approximated by Gaussian.

The corrections $\delta^{\rm HQ}_{\Lambda_c}(ij)$ appear along with the WCs that encode the NP contributions, as described in Eq.~\eqref{eq:delta_int}.
Table~\ref{Tab:scenario} summarizes the WCs for several NP scenarios motivated by the current $R_{D^{(*)}}$ anomaly.
We consider three ``single operator'' scenarios and three ``single leptoquark (LQ)'' scenarios.
The WCs are obtained from a global fit to the experimental data on $R_{D^{(*)}}$ and the $D^*$ longitudinal polarization $F_L^{D^*}$, as detailed in Ref.~\cite{Iguro:2024hyk}.
Although the fit is performed using HQET-based form factors, the following conclusions remain valid even when we adopt the BGL parameterization.
In the right panel of Fig.~\ref{fig:delta_HQ}, we show the correction to the sum rule prediction for $R_{\Lambda_c}$ in those scenarios.
Here, we express the deviation as\footnote{
In the last equation, since the Wilson coefficients are extracted, $\delta R_{\Lambda_c}^{kl}$ seems to depend on a scale $\mu$. 
However, this is not the case, because $\delta_{\Lambda_c}^{kl}$ in the middle expression is $\mu$-independent. 
}
\begin{align}
 \delta R_{\Lambda_c}^{kl} \equiv \delta_{\Lambda_c}^{kl} R^{\rm SM}_{\Lambda_c} = \sum_{ij} \mathcal{C}_i \mathcal{C}_j^* \,\delta_{\Lambda_c}^{kl}(ij) \, R^{\rm SM}_{\Lambda_c} \,.
\end{align}
Compared with the current experimental uncertainties in $R_{\Lambda_c}$, we find that the corrections are negligible in all scenarios.
Moreover, they are expected to remain smaller than the projected LHCb uncertainties except in the $T$, $\text{R}_2$, and $\text{S}_1$ scenarios when using the BGL parameterization for the $B \to D^{(*)}$ transitions. 
In these exceptional cases, NP contributions to the tensor operator $C_T$ become significant.
Therefore, further improvements in the determination of tensor form factors are essential for enabling reliable predictions from the sum rule.

\begin{figure}[!t]
\begin{center}
\includegraphics[width=0.41\linewidth]{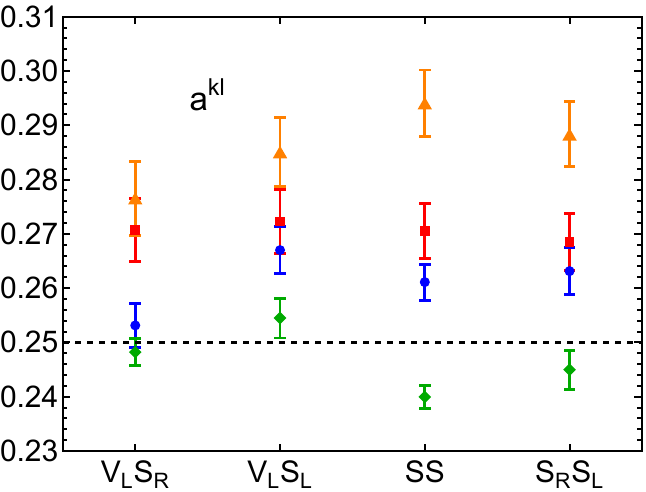}~~
\includegraphics[width=0.41\linewidth]{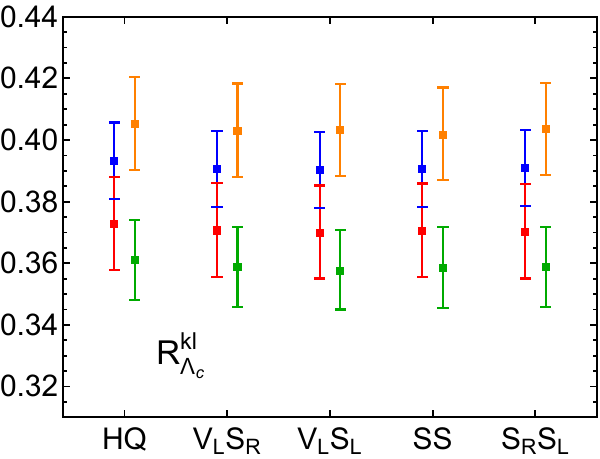}
\end{center}
\vspace{-.25cm}
\caption{
(Left) the sum rule coefficient $a^{kl}$ for various $\{kl\}$ and form factor parameterizations.
The coefficient $b^{kl}$ satisfies $a^{kl} + b^{kl} = 1$.
The horizontal dashed line $a^{kl}=1/4$ corresponds to the heavy quark and zero-recoil limits.
(Right) the sum rule predictions for $R_{\Lambda_c}$ with $68\,\%$ probability intervals are shown.
Here, the current experimental values of $R_D$ and $R_{D^*}$ are used. 
The correction $\delta^{kl}_{\Lambda_c}$ is assumed to be negligible. 
The color legend is the same as that in Fig.~\ref{fig:RLambda}.
}
\label{fig:kl-dep}
\end{figure}

As explained in Sec.~\ref{sec:formula_total}, some of $\delta_{\Lambda_c}^{kl}(ij)$ may be suppressed by appropriately shifting the sum rule coefficients.
In Eqs.~\eqref{eq:delta_int_mod} and \eqref{eq:coefficient_int_mod}, we choose $a^{kl}$ and $b^{kl}$ such that $\delta_{\Lambda_c}^{kl}(kl) = 0$ is satisfied.
In the left panel of Fig.~\ref{fig:kl-dep}, we show $a^{kl}$ for $\{kl\} = \{V_LS_R\}$, $\{V_LS_L\}$, $\{SS\}$, and $\{S_RS_L\}$, using various form factor parameterizations.
Also, in the right panel, we present the sum rule prediction for $R_{\Lambda_c}$ with $68\,\%$ probability intervals, given the current experimental values of $R_D$ and $R_{D^*}$. 
We find that, although $a^{kl}$ depends on the choice of $\{kl\}$, the predicted value of $R_{\Lambda_c}$ remains largely insensitive to this choice.

\begin{figure}[!t]
\begin{center}
\includegraphics[width=0.37\linewidth]{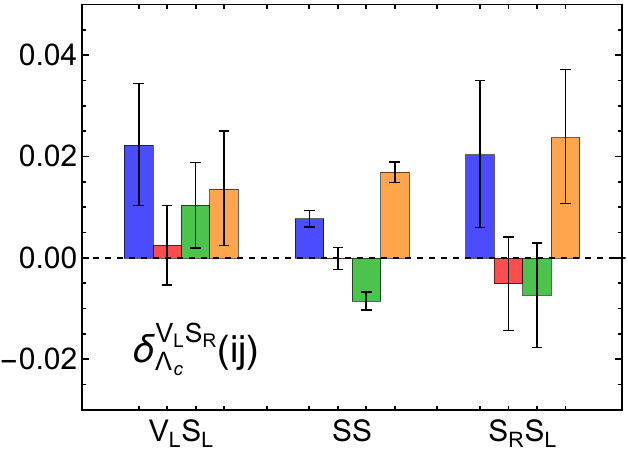}~
\includegraphics[width=0.227\linewidth]{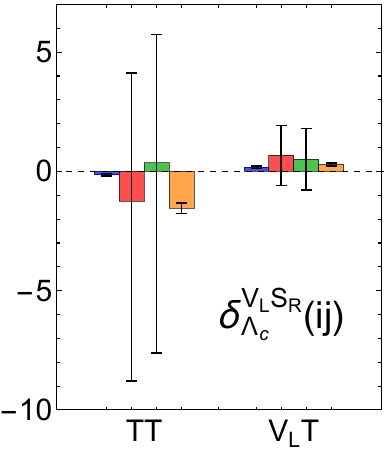}~
\includegraphics[width=0.365\linewidth]{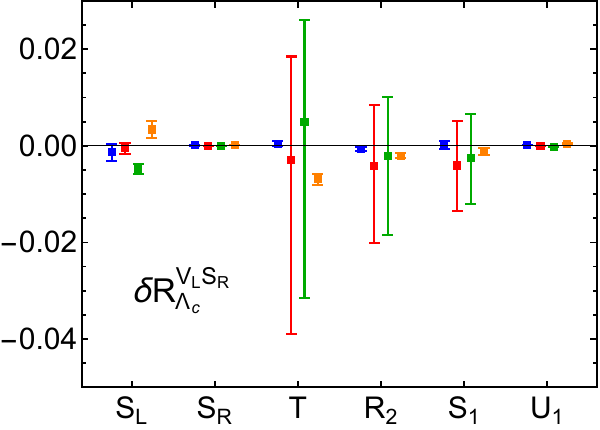}
\end{center}
\vspace{-.25cm}
\caption{
Same as Fig.~\ref{fig:delta_HQ} but for $\{kl\}=\{V_LS_R\}$.\vspace{.15cm}\\
}
\label{fig:delta_VSR}
\end{figure}

In Fig.~\ref{fig:delta_VSR}, we evaluate $\delta^{kl}_{\Lambda_c}$, focusing on the case of $\{kl\} = \{V_LS_R\}$. 
Results for other cases are provided in Appendix~\ref{sec:figures}.
Compared to the results in Fig.~\ref{fig:delta_HQ}, $\delta^{V_LS_R}_{\Lambda_c}(V_LS_R)$ vanishes, as expected. 
We find that the overall conclusion remains unchanged from Fig.~\ref{fig:delta_HQ}.
The uncertainties are below the current experimental value and the projected LHCb sensitivities, except in the $T$, $\text{R}_2$, and $\text{S}_1$ scenarios when the BGL parameterization is used for the $B \to D^{(*)}$ form factors.
The correction associated with the tensor operator exhibits large uncertainties.

\section{Conclusion and discussion}
\label{sec:conclusion}

In this paper, we examined the $b \to c$ semileptonic sum rule that relates the decay rates of $\Lambda_b \to \Lambda_c\tau\bar\nu$ to those for $B \to D^{(*)}\tau\bar\nu$.
The sum rule provides a non-trivial test of both experimental data and SM predictions. 
We outlined the derivation of the sum rule for the total decay rates, starting from the relation for the differential decay rates that holds exactly in the heavy quark limit.
By additionally considering the zero-recoil limit, we derived the sum rule of Eq.~\eqref{eq:RSR_limit}, where the coefficients are obtained as $a^{\rm HQ} = 1/4$ and $b^{\rm HQ} = 3/4$, with no correction term $\delta_R$.
We then identified the practical sources of $\delta_R$. 
We evaluated its size using both the HQET and BGL parameterizations. We also constructed probability distributions for the sum rule parameters through toy Monte Carlo simulations.

We showed that the sum rule is consistent with the current experimental results for $R_D$, $R_{D^*}$, and $R_{\Lambda_c}$.
Also, the correction $\delta_R$ is negligible compared to the uncertainties.
Accordingly, no inconsistencies are observed in the current experimental data and the SM predictions. 
Although some of $\delta^{kl}_{\Lambda_c}(ij)$ can be suppressed by appropriately shifting the sum rule coefficients $a^{kl}$ and $b^{kl}$, the conclusions drawn from Eq.~\eqref{eq:RSR_limit} are stable against such adjustment. 
Namely, the results remain unchanged from those with $a^{kl} = 1/4$ and $b^{kl} = 3/4$. 

In the future, experimental uncertainties in the measurement of $R_{\Lambda_c}$ are expected to decrease significantly at LHCb and Tera-Z.
For the sum rule to remain reliably applicable under such improvements, the form factor uncertainties particularly associated with the tensor operator need to be further reduced.

By comparing various form factor parameterizations, we found discrepancies among the sum rule predictions for $R_{\Lambda_c}$, which can be attributed to potential systematic uncertainties in the form factors. 
These discrepancies will become more significant in the future as $R_D$ and $R_{D^*}$ will be measured with greater precision at Belle~II.  
Therefore, further refinements in form factor determination are essential for making reliable predictions.

Finally, we would like to point out that the sum rule applies not only to the lepton-flavor universality ratios, but also to the total decay rates themselves.
In the heavy quark and zero-recoil limits, we derived the relation of Eq.~\eqref{eq:gamma_1} among $\Gamma_{D}$, $\Gamma_{D^*}$, and $\Gamma_{\Lambda_c}$.
Although these measurements do not benefit from cancellations of systematic uncertainties, the sum rule may still allow for a cross-check of the consistency of the experimental data in the future.
Besides, with large experimental statistics, the differential decay rates will be measured more precisely.
The sum rules among differential decay rates are discussed in Appendix~\ref{sec:SR_differential}.

\section*{Acknowledgements}
This work is supported by JSPS KAKENHI Grant Numbers 22K21347 [M.E. and S.I.], 24K07025 [S.M.], 24K22879 [S.I.], 24K23939 [S.I.] and 25K17385 [S.I.]. 
The work is also supported by JPJSCCA20200002 and the Toyoaki scholarship foundation [S.I.].
We also appreciate KEK-KMI joint appointment program [M.E. and S.I.], which accelerated this project. 
S.I. appreciates University of Z\"urich for the hospitality where he stayed at the last stage of the project.
\appendix

\section{Additional figures}
\label{sec:figures}

\begin{figure}[b]
\begin{center}
\includegraphics[width=0.37\linewidth]{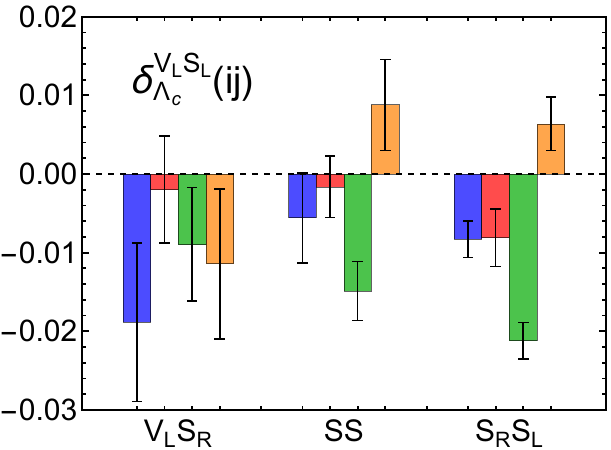}~
\includegraphics[width=0.227\linewidth]{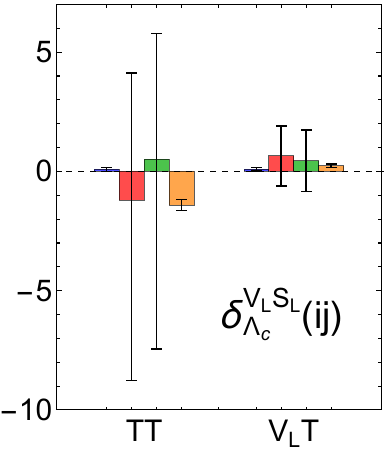}~
\includegraphics[width=0.365\linewidth]{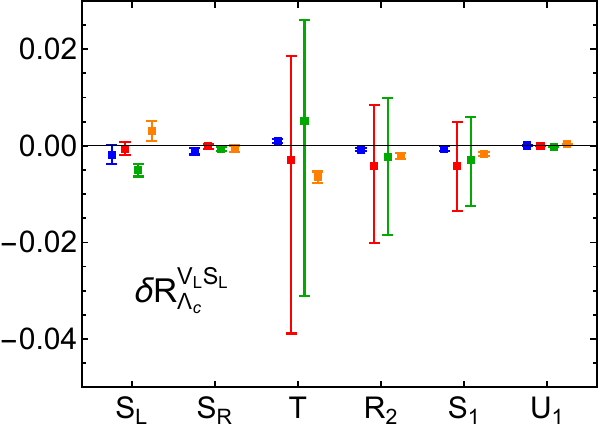}
\end{center}
\vspace{-.25cm}
\caption{
Same as Fig.~\ref{fig:delta_HQ} but for $\{kl\}=\{V_LS_L\}$.\vspace{.25cm}\\
}
\label{fig:delta_VSL}
\end{figure}

\begin{figure}[!t]
\begin{center}
\includegraphics[width=0.37\linewidth]{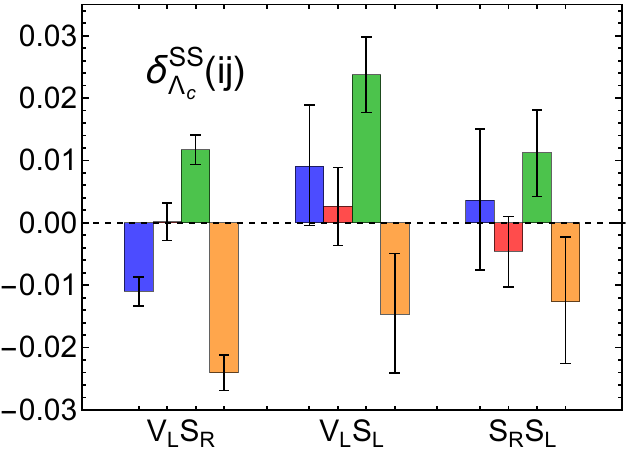}~
\includegraphics[width=0.227\linewidth]{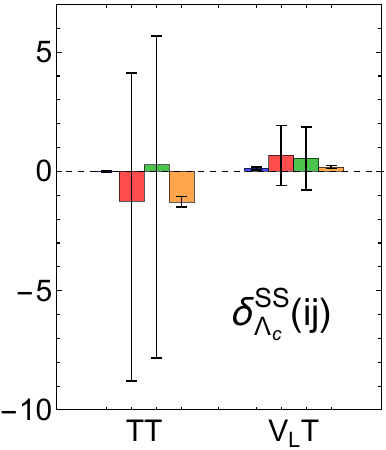}~
\includegraphics[width=0.365\linewidth]{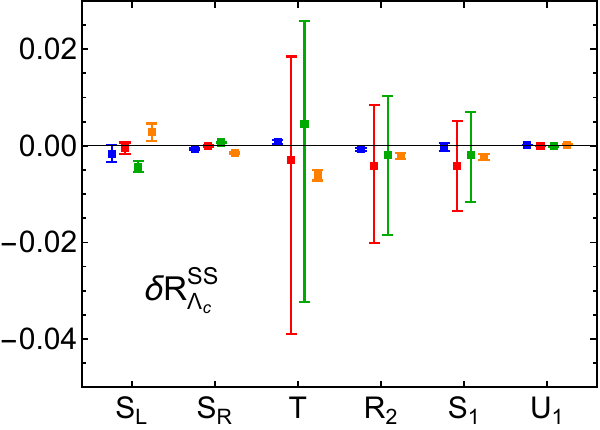}
\end{center}
\vspace{-.25cm}
\caption{
Same as Fig.~\ref{fig:delta_HQ} but for $\{kl\}=\{SS\}$.\vspace{.5cm}\\
}
\label{fig:delta_SS}
\end{figure}

\begin{figure}[!t]
\begin{center}
\includegraphics[width=0.37\linewidth]{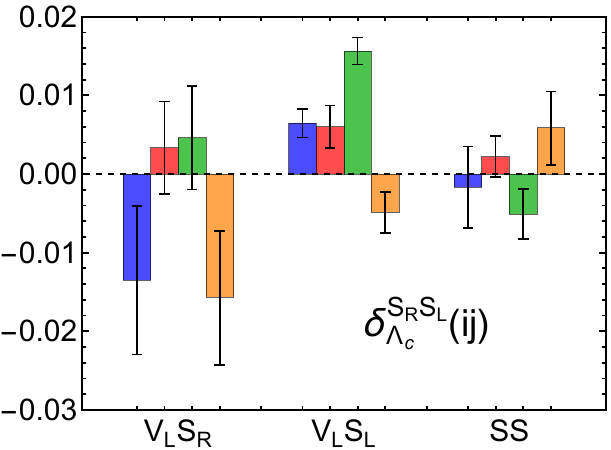}~
\includegraphics[width=0.227\linewidth]{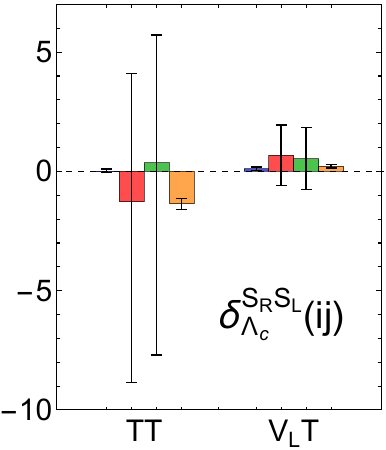}~
\includegraphics[width=0.365\linewidth]{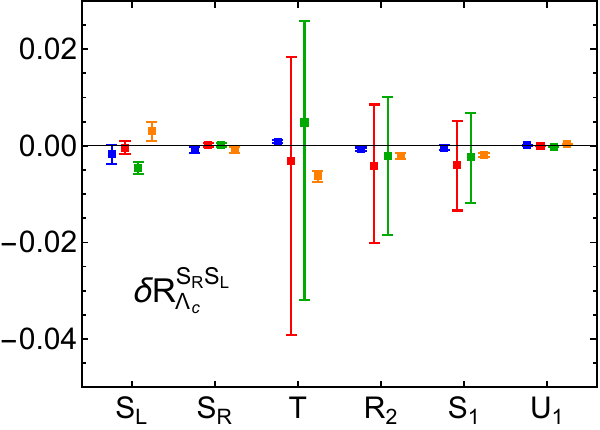}
\end{center}
\vspace{-.25cm}
\caption{
Same as Fig.~\ref{fig:delta_HQ} but for $\{kl\}=\{S_RS_L\}$.
}
\label{fig:delta_SLSR}
\end{figure}

In Eq.~\eqref{eq:delta_int_mod}, there are several choices of $\{kl\}$ that satisfy $\delta_{\Lambda_c}^{kl}(kl) = 0$.
In Fig.~\ref{fig:delta_VSR}, we considered $\{kl\} = \{V_LS_R\}$. 
In this appendix, we examine other choices.
Figures~\ref{fig:delta_VSL}, \ref{fig:delta_SS}, and \ref{fig:delta_SLSR} show the results for $\{kl\} = \{V_L S_L\}$, $\{SS\}$, and $\{S_L S_R\}$, respectively.
Although $\delta_{\Lambda_c}^{kl}$ depends on the choice of $\{kl\}$, the conclusion remains unchanged from Sec.~\ref{sec:result_int}.
The values of $\delta_{\Lambda_c}^{kl}$ are sufficiently small compared to both current experimental uncertainties and future projections, except for the cases involving the tensor operator.
It should be noted that, for $\{kl\} = \{TT\}$, the correction $\delta^{TT}_{\Lambda_c}(ij)$ for $(ij) \neq (TT)$ suffers from huge uncertainties originating from $\delta_{\Lambda_c}^{\rm HQ}(TT)$, thereby reducing the predictability for the sum rule.
The same argument applies to the case for $\{kl\} = \{V_LT\}$.

\section{Sum rule for differential decay rate}
\label{sec:SR_differential}

\subsection{Formulation}
\label{sec:formula_diff}

We can construct the sum rule for the differential decay rates starting from Eq.~\eqref{eq:DDRSR}, which was originally shown in Ref.~\cite{Endo:2025cvu}. 
Analogously to Eq.~\eqref{eq:gamma_4}, the sum rule can be obtained by normalizing the differential decay rates with the SM predictions as
\begin{align}
 \frac{\kappa_{\Lambda_c}}{\kappa_{\Lambda_c}^{\rm SM}} = a_{\kappa}^{\rm HQ} \frac{\kappa_{D}}{\kappa_{D}^{\rm SM}} + b_{\kappa}^{\rm HQ} \frac{\kappa_{D^*}}{\kappa_{D^*}^{\rm SM}} \,,
 \label{eq:HQ_sum_rule}
\end{align}
where the coefficients are given by
\begin{align}
 \label{eq:ab_HQL}
 a_{\kappa}^{\rm HQ} &= \frac{2}{1+w} \frac{\zeta(w)^2}{\xi(w)^2} \frac{\kappa_{D}^{\rm SM}}{\kappa_{\Lambda_c}^{\rm SM}} = \frac{(2+\rho^2)(1+r)^2(w-1) + 3\rho^2(1-r)^2(w+1)}{4(1+2\rho^2)(r^2w+w-2r)+ 4(2+\rho^2) w \hat q^2} \,, \\
 b_{\kappa}^{\rm HQ} &= \frac{2}{1+w} \frac{\zeta(w)^2}{\xi(w)^2} \frac{\kappa_{D^{*}}^{\rm SM}}{\kappa_{\Lambda_c}^{\rm SM}} = \frac{(2+\rho^2) \Big[ (1-r)^2(w+1) + 4w\hat q^2 \Big] + 3\rho^2(1+r)^2(w-1)}{4(1+2\rho^2)(r^2w+w-2r)+ 4(2+\rho^2) w \hat q^2} \,.
 \notag 
\end{align}
They are functions of $w$ and hadron masses, but are independent of NP contributions, $C_X$. 
They satisfy the relation, $a_{\kappa}^{\rm HQ} + b_{\kappa}^{\rm HQ} = 1$.
The SM value $\kappa_{c}^{\rm SM}$ is given by $C_{X} = 0$.

The above sum rule holds exactly in the heavy quark limit.
However, similar to Sec.~\ref{sec:SR_total}, corrections to the sum rule arise in practice, {\it i.e.,} by realistic hadron masses and higher-order contributions to the form factors.
They are expressed as 
\begin{align}
 \overline{\delta}_{\Lambda_c}^{\rm HQ} \equiv \sum_{ij}  \mathcal{C}_i \mathcal{C}_j^* \,\overline{\delta}_{\Lambda_c}^{\rm HQ}(ij)\,,
 ~~~
 \overline{\delta}_{\Lambda_c}^{\rm HQ}(ij) = 
 \frac{\kappa^{ij}_{\Lambda_c}}{\kappa^{\rm SM}_{\Lambda_c}} - a_{\kappa}^{\rm HQ} \frac{\kappa^{ij}_{D}}{\kappa_{D}^{\rm SM}} - b_{\kappa}^{\rm HQ} \frac{\kappa^{ij}_{D^*}}{\kappa^{\rm SM}_{D^*}} \,,
 \label{eq:delta_mod_1}
\end{align}
Furthermore, by modifying the coefficients, the correction can be redefined as
\begin{align}
 \overline{\delta}_{\Lambda_c}^{kl} \equiv \sum_{ij}  \mathcal{C}_i \mathcal{C}_j^* \,\overline{\delta}_{\Lambda_c}^{kl}(ij)\,,
 ~~~
 \overline{\delta}_{\Lambda_c}^{kl}(ij) = 
 \frac{\kappa^{ij}_{\Lambda_c}}{\kappa^{\rm SM}_{\Lambda_c}} - a_{\kappa}^{kl} \frac{\kappa^{ij}_{D}}{\kappa_{D}^{\rm SM}} - b_{\kappa}^{kl} \frac{\kappa^{ij}_{D^*}}{\kappa^{\rm SM}_{D^*}} \,.
 \label{eq:delta_mod_2}
\end{align}
Here, the coefficients are chosen such that $\overline{\delta}_{\Lambda_c}^{kl}(kl) = 0$ is satisfied, which is achieved by setting $a_{\kappa}^{kl}$ and $b_{\kappa}^{kl}$ as
\begin{align}
 a_{\kappa}^{kl} = \frac{\kappa_{\Lambda_c}^{kl}/\kappa_{\Lambda_c}^{V_LV_L}-\kappa_{D^*}^{kl}/\kappa_{D^*}^{V_LV_L}}{\kappa_{D}^{kl}/\kappa_{D}^{V_LV_L}-\kappa_{D^*}^{kl}/\kappa_{D^*}^{V_LV_L}} \,,~~~
 b_{\kappa}^{kl} = \frac{\kappa_{D}^{kl}/\kappa_{D}^{V_LV_L}-\kappa_{\Lambda_c}^{kl}/\kappa_{\Lambda_c}^{V_LV_L}}{\kappa_{D}^{kl}/\kappa_{D}^{V_LV_L}-\kappa_{D^*}^{kl}/\kappa_{D^*}^{V_LV_L}} \,.
 \label{eq:ab_mod}
\end{align}
Since the SM values are given by $\kappa_{H_c}^{\rm SM} = \kappa_{H_c}^{V_LV_L}$, the correction satisfies $\overline{\delta}_{\Lambda_c}^{kl}(V_LV_L) = 0$.

In Eq.~\eqref{eq:HQ_sum_rule}, the differential decay rates are not normalized with those of the light lepton channels. 
However, such a sum rule can be derived by applying the following replacement,
\begin{align}
 \frac{\kappa_{H_c}}{\kappa_{H_c}^{\rm SM}} \to \frac{\kappa_{H_c}/\kappa_{H_c}^\ell}{\kappa_{H_c}^{\rm SM}/\kappa_{H_c}^{\ell,\, {\rm SM}}}\,,
\end{align}
where the differential decay rate for the light-lepton channels is defined as $\kappa_{H_c}^\ell = d\Gamma (H_b \to H_c \ell\bar\nu_\ell)/dw$ with $\ell = e$, $\mu$.
Since we assume that NP contributes only to $b\to c \tau\bar\nu$, the relation $\kappa_{H_c}^\ell = \kappa_{H_c}^{\ell,\, {\rm SM}}$ holds.

\subsection{Numerical result}

\begin{figure}[!t]
\begin{center}
\includegraphics[width=0.35\linewidth]{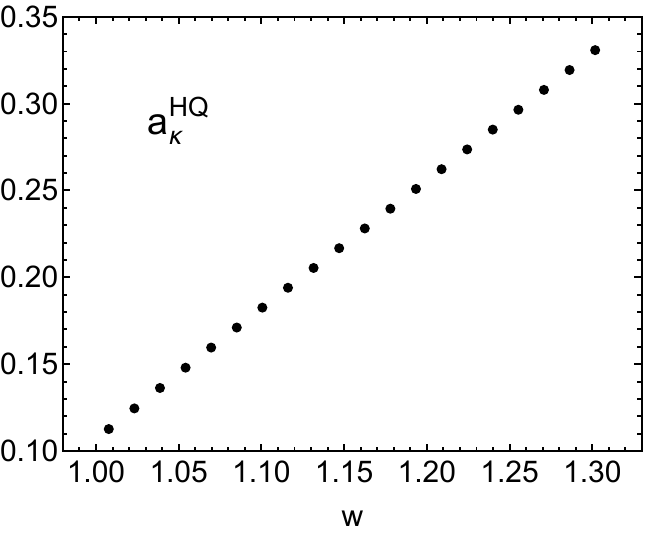}
\end{center}
\vspace{-.25cm}
\caption{
Coefficient of the sum rule for the differential decay rates, $a_{\kappa}^{\rm HQ}$, as a function of $w$.
Here, the hadron masses are set to be $m_B$ ($m_D$) for the bottomed (charmed) hadrons.
}
\label{fig:diff_coefficient_HQ}
\end{figure}

Similar to Sec.~\ref{sec:result_int}, we evaluate the correction to the sum rule for the differential decay rate.
We consider the HQET and BGL form factors.
Currently, no differential data is available for $\Lambda_b \to \Lambda_c \tau\bar\nu$.
Although experimental data is available for the differential decay rates of $B\to D^{(*)} \tau\bar\nu$, the uncertainty is still large \cite{BaBar:2013mob,Belle:2015qfa} and hence we will not predict $\Lambda_b \to \Lambda_c \tau\bar\nu$ from the existing $B\to D^{(*)} \tau\bar\nu$ data.

Let us begin by evaluating Eq.~\eqref{eq:delta_mod_1}, where the sum rule coefficients are given by Eq.~\eqref{eq:ab_HQL}.
The sum rule can be defined physically in the range of $1 < w < w_{\rm max}^\kappa$, where $w_{\rm max}^\kappa\simeq 1.3$ is the minimum value of $w_{\rm max}$ among $B\to D \tau\bar\nu$, $B\to D^{*} \tau\bar\nu$, and $\Lambda_b \to \Lambda_c \tau\bar\nu$.

In Fig.~\ref{fig:diff_coefficient_HQ}, we show the coefficient $a_{\kappa}^{\rm HQ}$ as a function of $w$.
Here, the hadron masses are set to be $m_B$ ($m_D$) for the bottomed (charmed) hadrons.
The coefficients satisfy $a_{\kappa}^{\rm HQ} + b_{\kappa}^{\rm HQ} = 1$.
Since the hadron masses have been determined precisely, we ignore uncertainties associated with them.

\begin{figure}[!t]
\begin{center}
\includegraphics[width=0.305\linewidth]{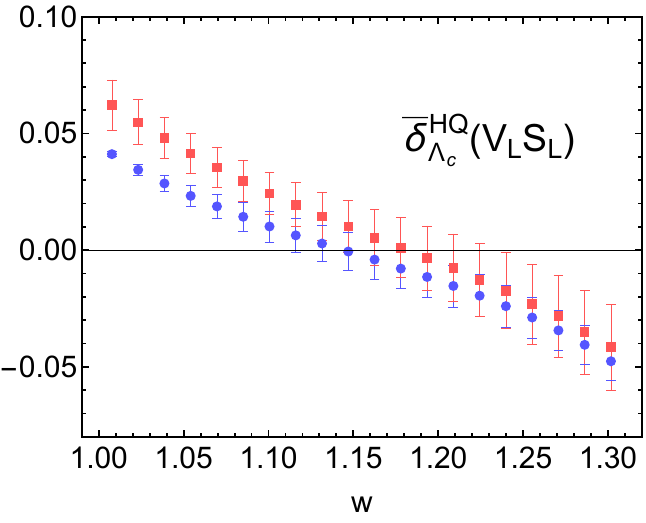}~~
\includegraphics[width=0.303\linewidth]{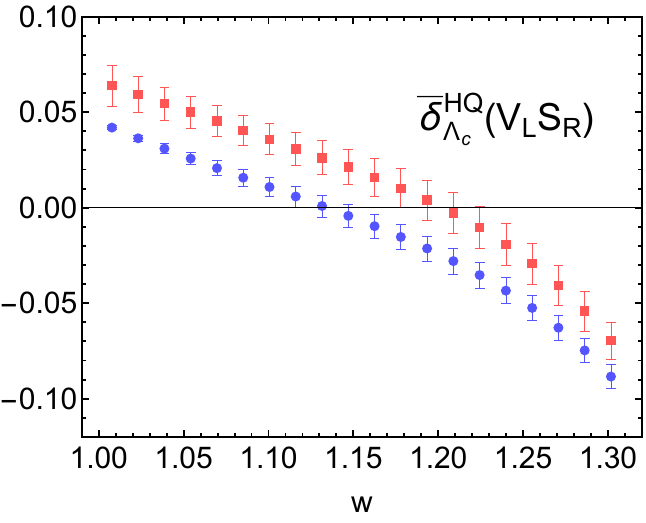}~~
\includegraphics[width=0.302\linewidth]{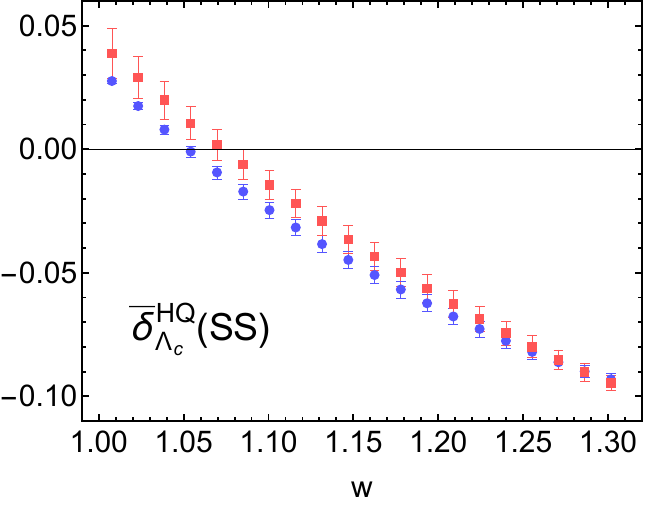} \vspace{0.2cm}\\
\includegraphics[width=0.305\linewidth]{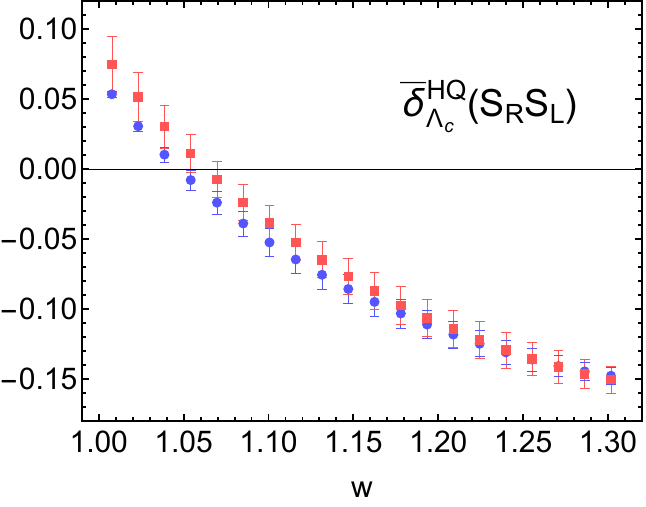}~~
\includegraphics[width=0.293\linewidth]{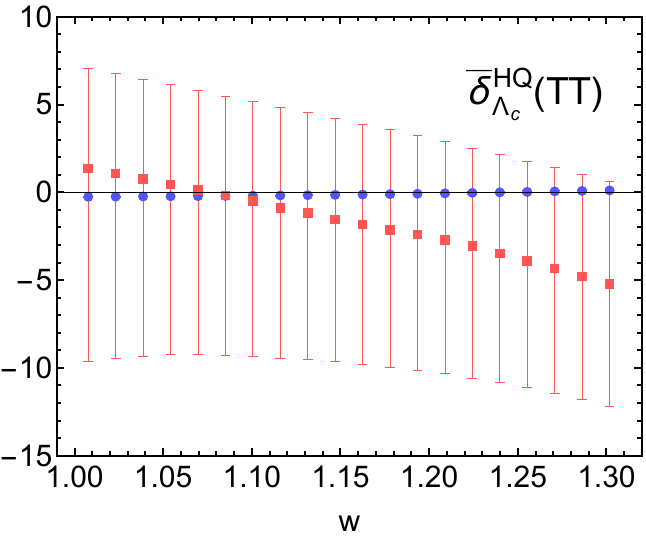}~~
\includegraphics[width=0.285\linewidth]{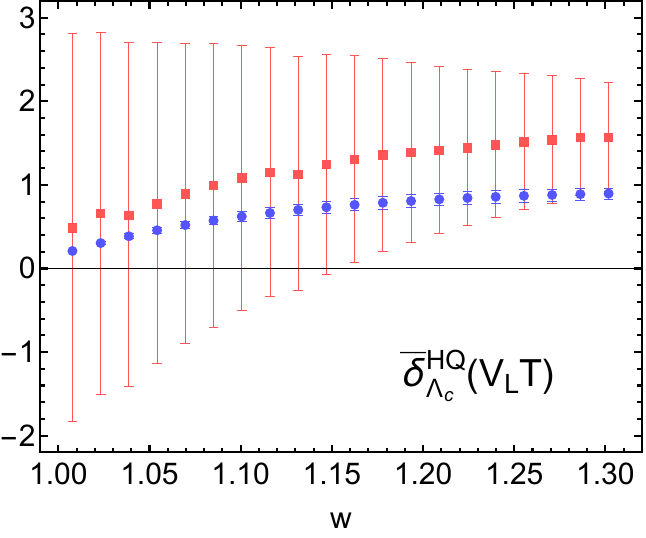}~~
\mbox{\raisebox{18mm}{\includegraphics[width=0.1\linewidth]{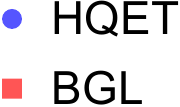}}}
\end{center}
\vspace{-.25cm}
\caption{
Corrections to the sum rule for the differential decay rates as a function of $w$.
The sum rule coefficients are given by Eq.~\eqref{eq:ab_HQL}, {\it i.e.}, those in the heavy quark limit, where the masses are set to be $m_B$ ($m_D$) for the bottomed (charmed) hadrons. \vspace{.25cm}\\
}
\label{fig:differential_HQL}
\end{figure}

In Fig.~\ref{fig:differential_HQL}, we show $\overline{\delta}_{\Lambda_c}^{\rm HQ}(ij)$ for $(ij) = (V_LS_L)$, $(V_LS_R)$, $(SS)$, $(S_RS_L)$, $(TT)$, and $(TV_L)$.
The sum rule coefficients are given by Eq.~\eqref{eq:ab_HQL}.
Here, the hadron masses are set as those in Fig.~\ref{fig:diff_coefficient_HQ}.
The form factors are evaluated using HQET and BGL.
We find that the corrections are generally suppressed, except for the cases $(ij) = (TT)$ and $(TV_L)$.
In particular, the corrections from the tensor operator exhibit much larger uncertainties associated with the form factors for the BGL parameterization, similar to the results in Sec.~\ref{sec:result_int}. 
We also observe discrepancies between the results based on HQET and those based on BGL, especially for $(ij) = (V_LS_L)$ and $(V_LS_R)$.
These discrepancies are considered to arise from potential systematic uncertainties in the form factors.
Therefore, they are required to be reduced by further determining the vector and scalar form factors using updated lattice QCD calculations and experimental data, complemented by theoretical inputs. 

\begin{figure}[!t]
\begin{center}
\includegraphics[width=0.35\linewidth]{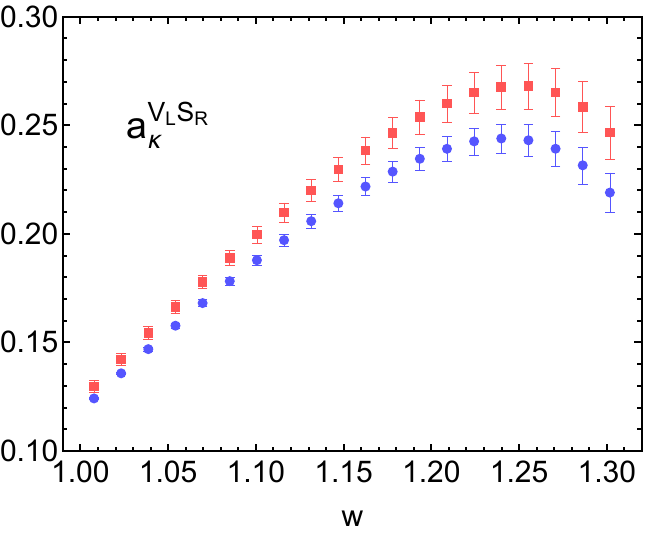}
\end{center}
\vspace{-.15cm}
\caption{
Coefficient of the sum rule for the differential decay rates, $a^{V_LS_R}_{\kappa}$. The coefficient $b^{V_LS_R}_{\kappa}$ satisfies $a^{V_LS_R}_{\kappa}+b^{V_LS_R}_{\kappa}=1$.\vspace{.15cm}\\
}
\label{fig:diff_coefficient_VSR}
\end{figure}

\begin{figure}[!t]
\begin{center}
\includegraphics[width=0.3\linewidth]{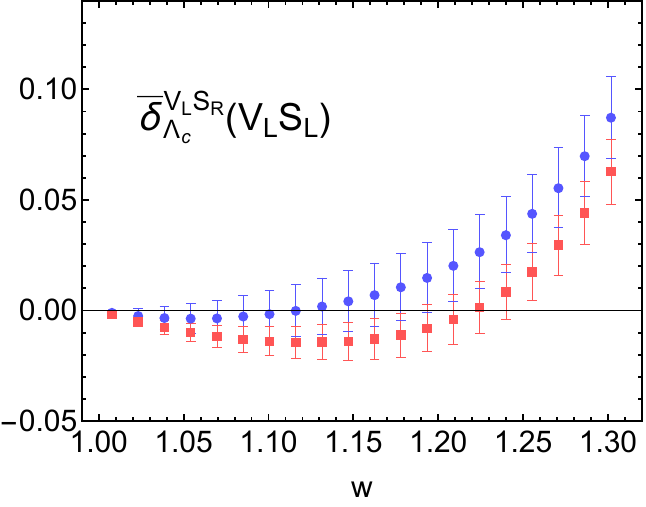}~~
\includegraphics[width=0.3\linewidth]{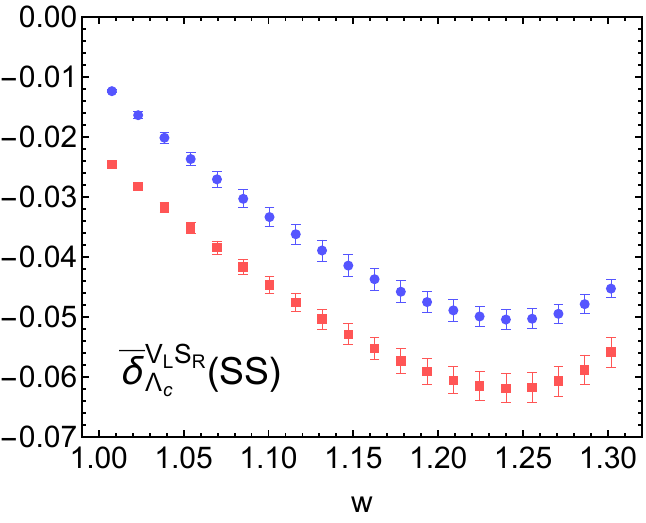}~~
\includegraphics[width=0.3\linewidth]{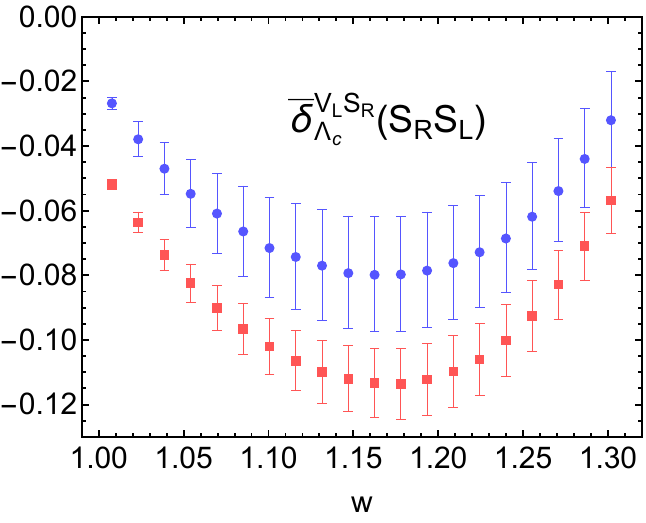} \vspace{0.2cm}\\
\includegraphics[width=0.3\linewidth]{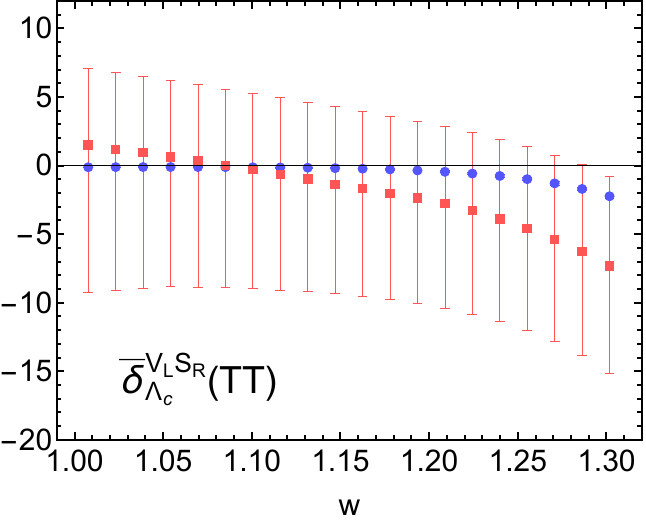}~~
\includegraphics[width=0.295\linewidth]{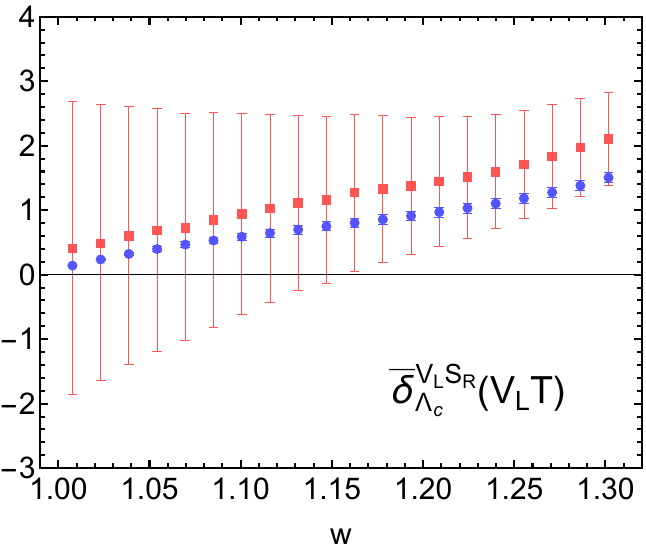}
\end{center}
\vspace{-.25cm}
\caption{
Corrections to the sum rule for the differential decay rates, $\overline{\delta}^{V_LS_R}_{\Lambda_c}(ij)$.
}
\label{fig:differential_VSR}
\end{figure}

Similar to Sec.~\ref{sec:result_int}, we consider the case in which some of $\overline{\delta}_{\Lambda_c}^{kl}(ij)$ are suppressed by appropriately choosing the sum rule coefficients. 
In Figs.~\ref{fig:diff_coefficient_VSR} and \ref{fig:differential_VSR}, we show the results for $\{kl\} = \{V_LS_R\}$ as a reference case. 
By comparing Fig.~\ref{fig:diff_coefficient_VSR} with Fig.~\ref{fig:diff_coefficient_HQ}, we see that the discrepancy of $a^{V_LS_R}_{\kappa}$ arising from the form factor parameterizations becomes enhanced as $w$ increases.
On the other hand, the correction $\overline{\delta}_{\Lambda_c}^{kl}(ij)$ depends on the choice of $\{kl\}$.
In particular, the discrepancy between the results based on HQET and those based on BGL becomes significant for $(ij) = (SS)$, which is originally attributed to the difference appearing in $\overline{\delta}_{\Lambda_c}^{\rm HQ}(V_LS_R)$.
Additional inputs for the form factors may alleviate this discrepancy and reduce the uncertainties associated with the tensor operator. 
In conclusion, the corrections to the sum rule exhibit discrepancies arising from the form factor parameterizations.
Although experimental results are not yet precise, we need to improve the form factor determinations.

\bibliographystyle{utphys28mod}
\bibliography{ref}
\end{document}